\documentclass[twocolumn,twocolappendix]{aastex701} 
\usepackage{graphicx}	
\usepackage{amsmath}	

\usepackage{longtable}
\usepackage{CJK}

\begin{document}
\begin{CJK*}{UTF8}{gbsn}

\title{Spectral Evidence of Heavy Nuclei from the Neutron Star Crust in Magnetar Bursts}
\shortauthors{Xie et al.}

\correspondingauthor{Yun-Wei Yu}
\email{yuyw@ccnu.edu.cn}
\correspondingauthor{Shao-Lin Xiong}
\email{xiongsl@ihep.ac.cn}

\author[0000-0001-9217-7070]{Sheng-Lun Xie (谢升伦)}
\affiliation{Institute of Astrophysics, Central China Normal University, Wuhan 430079, China}
\affiliation{State Key Laboratory of Particle Astrophysics, Institute of High Energy Physics, Chinese Academy of Sciences, 19B Yuquan Road, Beijing 100049, China}
\email{xiesl@mails.ccnu.edu.cn}
\author[0000-0002-1067-1911]{Yun-Wei Yu (俞云伟)}
\affiliation{Institute of Astrophysics, Central China Normal University, Wuhan 430079, China}
\email{yuyw@ccnu.edu.cn}
\author[0000-0002-4771-7653]{Shao-Lin Xiong (熊少林)}
\affiliation{State Key Laboratory of Particle Astrophysics, Institute of High Energy Physics, Chinese Academy of Sciences, 19B Yuquan Road, Beijing 100049, China}
\email{xiongsl@ihep.ac.cn}
\author[0000-0001-8664-5085]{Wang-Chen Xue (薛王陈)}
\affiliation{State Key Laboratory of Particle Astrophysics, Institute of High Energy Physics, Chinese Academy of Sciences, 19B Yuquan Road, Beijing 100049, China}
\affiliation{University of Chinese Academy of Sciences, Beijing 100049, China}
\email{xuewc@ihep.ac.cn}

\begin{abstract}
The crust of a neutron star (NS) provides a unique laboratory for studying matter under extreme density and magnetic field conditions that cannot be realized in terrestrial experiments.
However, direct observational constraints on its composition have remained very limited.
Magnetar bursts provide a promising means to probe the nuclear composition of the outer crust, as their energy release may be associated with stress-driven yielding of the crustal Coulomb lattice (including plastic deformation) and magnetic reconnection in the surrounding magnetosphere.
We develop a general-purpose radiative transfer framework for a strongly magnetized electron--ion thermal plasma (MEITP) and apply it to the observed X-ray burst spectra.
The spectral fits disfavor light-ion compositions and instead favor plasmas characterized by effective charge numbers around $Z \sim 37$.
These results provide spectral evidence for the participation of heavy nuclei in magnetar bursts, offer new observational constraints on the baryonic content and the location of the emitting fireballs, and further imply a crustal origin of the heavy ions.
\end{abstract}


\section{Introduction}\label{sec:intro}
Magnetars are a class of isolated neutron stars (NSs) characterized by extremely strong dipole magnetic fields \citep[$\sim10^{14}$--$10^{15}$~G;][]{Duncan1992ApJ,Thompson1993ApJ}, and are generally identified with soft $\gamma$-ray repeaters (SGRs) and anomalous X-ray pulsars \citep[AXPs;][]{Mereghetti2008AARv}. 
They exhibit a wide variety of high-energy transient burst activities spanning a broad range of energetics and timescales. 
Most events are short bursts lasting from milliseconds to $\lesssim 1$~s, with typical total energies up to $\sim10^{41}$~erg \citep[e.g.,][]{Woods2006csxs,Mereghetti2008AARv}.
More energetic events, often referred to as intermediate flares, last from seconds to tens of seconds and release $\sim10^{41}$--$10^{43}$~erg in total \citep[e.g.,][]{Olive2004ApJ,Woods2006csxs,Mereghetti2008AARv}.
The empirical distinction between short bursts and intermediate flares is not always sharp, and recent work has also provided new insights based on burst temporal morphologies \citep[e.g., ERCOD shape;][]{Wang2025ApJ}. 
Rarely, magnetars produce giant flares with total energies reaching $\sim10^{44}$--$10^{46}$~erg \citep[e.g.,][]{Mazets1979Natur,Woods2006csxs,Mereghetti2008AARv}.

The physical origin of magnetar bursts is commonly discussed in terms of rapid magnetic energy release associated with crustal activity \citep[i.e., starquakes, e.g.,][]{Jones2003ApJ,Levin2012MNRAS,Lander2015MNRAS,Dehman2020ApJ} and/or magnetospheric reconnection \citep[e.g.,][]{Duncan1992ApJ,Paczynski1992AcA,Masada2010PASJ,Yu2011ApJ,Yu2013ApJ,Meng2014ApJ}. 
Despite the diversity of proposed triggering mechanisms, it is widely accepted that the released energy rapidly forms an optically thick plasma fireball magnetically trapped near the magnetar surface \citep[e.g.,][]{Duncan1992ApJ,Paczynski1992AcA}, which mediates the observed high-energy radiation \citep[e.g.,][]{Thompson1995MNRAS,Thompson1996ApJ}.

From an observational perspective, the spectra of magnetar bursts are often described using empirical models that provide statistically acceptable fits. However, the inferred parameters are typically limited to phenomenological quantities, such as characteristic temperatures of order $\sim 10~\rm keV$, spectral peaks of order $E_{\rm peak}\sim30~\rm keV$, and simple spectral shape parameters with photon indices of $\sim -0.5$, and do not directly constrain the plasma density, magnetic field strength, baryon loading, or ionic composition \citep[e.g.,][]{Gogus1999ApJL,Lin2011ApJ,vonKienlin2012ApJ,Rehan2025ApJS}.
As a result, the ionic component is frequently not treated as an explicit parameter in burst spectral modeling but instead neglected or treated with simplified assumptions, for example, by adopting a purely pair-dominated plasma or restricting the composition to light ions \citep{Thompson1995MNRAS,Lyubarsky2002MNRAS,Wada2023MNRAS,Zhang2023MNRAS}.

If a substantial fraction of the burst energy is released near the magnetar surface, crustal material may be entrained into the fireball during its formation and subsequently flow outward along open magnetic field lines, leading to baryon loading \citep[e.g.,][]{Thompson2001ApJ,vanPutten2016MNRAS,Ioka2020ApJ,Yang2021ApJ,Wada2023MNRAS}.
The energy available from crustal failure alone can already reach the range relevant for bright bursts or intermediate flares \citep[$\gtrsim10^{40}$--$10^{42}$~erg;][]{Baiko2018MNRAS,Pons2011ApJ,Perna2011ApJ,Dehman2020ApJ,Zhang2022ApJ}.
These considerations suggest that a nonnegligible amount of crustal ions may be injected into the fireball, potentially affecting its radiative properties.

These considerations motivate a direct observational test of whether ions from the crust participate in the emitting plasma and how their presence influences radiative transfer. 
In this work, we develop a general-purpose radiative transfer framework for a strongly magnetized electron--ion thermal plasma (MEITP) and apply it to the observed spectra of bright magnetar bursts to constrain the local magnetic field, plasma conditions, and the effective ionic charge.

The rest of this paper is organized as follows: In Section \ref{sec:radiation_process}, we model the radiation of  the MEITP model.
Then, we implement the model to the observed magnetar bursts, as shown in Section \ref{sec:data_fit}.
Finally, we discuss the baryonic composition in the NS crust and the probable plasma fireball scenario in Section \ref{sec:discussion}.

\section{Radiation process}\label{sec:radiation_process}
Magnetar burst spectra are often described phenomenologically using empirical prescriptions (e.g., a power law with an exponential cutoff), which can provide acceptable fits but do not uniquely map to the underlying plasma conditions \citep{Lin2011ApJ,vonKienlin2012ApJ}. 
Alternative multicomponent thermal descriptions (e.g., two- or even three-blackbody models) have also been adopted \citep[e.g.,][]{Olive2004ApJ}, yet their physical interpretation in terms of a concrete radiation mechanism remains degenerate. 
Moreover, recent analyses have shown that fitting bright burst spectra with a single blackbody can leave systematic residuals \citep{Xie2026ApJ}, suggesting that thermal emission and radiative transfer in a strongly magnetized environment require a more physically motivated treatment. 
Motivated by these considerations, we adopt a general MEITP framework with reduced prior assumptions and summarize the key radiation processes below.

Given that the observed photon energy of typical magnetar bursts is in the range of $\lesssim 100$ keV \citep[e.g.,][]{Gogus1999ApJL,Gogus2000ApJL,Gavriil2004ApJ}, we use the low-energy limit of Compton scattering $\omega \ll m_{\mathrm{e}}c^2/\hbar$ to approximate the photon scattering in this work.
The thermal spectrum of the medium (e.g., magnetized plasmas or fireball bubbles) is characterized by free--free emission and absorption, as well as by saturated inverse Compton scattering.
As noted in \cite{RybickiLightman1986book}, at low frequencies, the spectrum is characterized as a blackbody, which then transitions to a modified blackbody spectrum \citep[e.g.,][]{FeltenRees1972AA,Illarionov1972SvA,RybickiLightman1986book}, varying the fraction of absorption and scattering across different frequencies. At high frequencies, the spectrum evolves into a Wien spectrum.
The Compton scattering effects in different frequencies are as follows.

\subsection{Coherent scattering effects}\label{subsec:coherent_scatter}
One may assess the change in energy of a photon as it traverses a finite medium by defining a Compton parameter $y$ as \citep{RybickiLightman1986book}
\begin{equation}
    y\equiv\begin{pmatrix}\text{number of}\\\text{scatterings}\end{pmatrix}\times\begin{pmatrix}\text{energy change}\\\text{per scattering}\end{pmatrix}.
\end{equation}
For $y\lesssim1$, coherent scattering is more significant than inverse Compton (incoherent scattering), as the energy change during scattering is relatively small. Conversely, inverse Compton becomes more important when $y\gtrsim1$ (see Section \ref{subsec:inverse_compt}).
Therefore, the emergent intensity in a scattering and absorbing medium may be evaluated as \citep{RybickiLightman1986book},
\begin{equation}\label{eq:Iv}
\begin{aligned}
    I_{\nu}^{\mathrm{E}}&=\frac{2B_\nu\sqrt{\epsilon}}{1+\sqrt{\epsilon}}, \\
    B_\nu(T)&=\frac2{c^2}\frac{h\nu^3}{e^{h\nu/kT}-1},
\end{aligned}
\end{equation}
where $B_\nu$ is the blackbody intensity and $\epsilon$ is the probability per interaction that the photon will be absorbed.
In the optically thick case, Eq. \ref{eq:Iv} can be applied to the entire spectrum, and if $\epsilon\sim1$, that equation will reduce to a blackbody spectrum, as absorption is more significant than scattering at low frequencies \citep{RybickiLightman1986book}.

The photon absorbing probability $\epsilon$ is estimated by
\begin{equation}
    \epsilon=\frac{\kappa_{\mathrm{ff}}}{\kappa_{\mathrm{es}}+\kappa_{\mathrm{ff}}},
\end{equation}
where $\kappa_{\mathrm{ff}}$ and $\kappa_{\mathrm{es}}$ are the electron free--free absorption and scattering coefficients, respectively.
For a plasma in a strongly magnetized environment, the absorption and scattering coefficients need to be reevaluated, as the effects of vacuum polarization on the spectrum become significant when the magnetic field strength is approximately the quantum critical field $B_\mathrm{Q}=m_\mathrm{e}^2c^3/(\hbar e)=4.4\times10^{13}$ G.
The particle $s$ scattering opacity coefficients are given by \citep[e.g.,][]{Ventura1979PhRvD,VenturaNagel1979ApJ,Nagel1980ApJ,HoLai2003MNRAS}
\begin{equation}\label{eq:kap_scat}
\begin{aligned}
\kappa_j^{\mathrm{es,s}}&=n_\mathrm{s}\sigma_\mathrm{T}\sum_{\alpha=-1}^1\frac{\omega^2}{(\omega-\alpha\omega_\mathrm{Bs})^2+\nu_\mathrm{s}^2}\left|e_\alpha^j\right|^2A_\alpha\\
&\times
\begin{cases}
1, &s=e\:(\mathrm{i.e.,\: electron})\\
\left(Z^2m_\mathrm{e}/Am_\mathrm{p}\right)^2, &s=i\:(\mathrm{i.e.,\: ion})
\end{cases}
\end{aligned}
\end{equation}
where $j$=1, 2 for the X-mode and O-mode
\footnote{The two photon polarization modes are termed the ordinary mode (O-mode, j=2), which is polarized parallel to the $\vec{k}-\vec{B}$ plane, where $\vec{k}$ is the photon wavevector and $\vec{B}$ is the magnetic field, and the extraordinary mode (X-mode, j=1), which is polarized perpendicular to that plane.}.
$Z$ and $A$ are the charge and the mass number of ion, respectively. 
$\sigma_\mathrm{T}$ is the Thomson cross section, $m_\mathrm{e}$ is the electron mass, $n_\mathrm{s}$ is the number density of the particle, $\omega_\mathrm{Bs}$ is the cyclotron frequency of the particle in the plasma, and $\nu_\mathrm{s}$ is the damping rate of the particle.
The cyclic component $\left|e_\alpha^j\right|$ is given by
\begin{equation}\label{eq:cyc_comp}
\begin{aligned}
|e_\pm^j|^2&=\frac{\left[1\pm\left(K_j\cos\theta+K_{z,j}\sin\theta\right)\right]^2}{2\left(1+K_j^2+K_{z,j}^2\right)} \\
\left|e_z^j\right|^2&=\frac{(K_j\sin\theta-K_{z,j}\cos\theta)^2}{1+K_j^2+K_{z,j}^2},
\end{aligned}
\end{equation}
where subscripts $\pm$ and $z$ specify the $\pm1$ and 0, respectively.
$\theta$ is the angle between the photon wavevector $\vec{k}$ and the magnetic field $\vec{B}$. For detailed calculations on the angle $A_\alpha = \sum_{i=1}^2A_\alpha^i$ and the ellipticity $K_j$ and $K_{z,j}$, see Appendix \ref{apx_sec:opacity}.

The opacity coefficients of free--free absorption are given by \citep[e.g.,][]{Virtamo1975NCimB,Pavlov1976JETP,Nagel1980ApJ,Meszaros1992book,HoLai2003MNRAS},
\begin{equation}\label{eq:kap_abs}
\begin{aligned}
\kappa_j^{\mathrm{ff,s}}&=\alpha_0\sum_{\alpha=-1}^1\frac{\omega^2}{(\omega-\alpha\omega_\mathrm{Bs})^2+\nu_\mathrm{s}^2}\left|e_\alpha^j\right|^2g_\alpha^{\mathrm{ff}}\\
&\times
\begin{cases}
1, &s=e\:(\mathrm{i.e.,\: electron})\\
\left(Z^2m_\mathrm{e}/Am_\mathrm{p}\right)^2/Z^3, &s=i\:(\mathrm{i.e.,\: ion})
\end{cases}
\end{aligned}
\end{equation}
where $\alpha_0$ and $g_\alpha^{\mathrm{ff}}$ are the free--free absorption coefficient and free--free Gaunt factor (see Appendix \ref{apx_sec:opacity}).

The total scattering or absorption opacity is then estimated as the sum of the contributions from electrons and ions, as well as the X- and O-mode components,
$\kappa^{\mathrm{es}}=\kappa_{1}^{\mathrm{es,e}}+\kappa_{1}^{\mathrm{es,i}}+\kappa_{2}^{\mathrm{es,e}}+\kappa_{2}^{\mathrm{es,i}}$ and $\kappa^{\mathrm{ff}}=\kappa_{1}^{\mathrm{ff,e}}+\kappa_{1}^{\mathrm{ff,i}}+\kappa_{2}^{\mathrm{ff,e}}+\kappa_{2}^{\mathrm{ff,i}}$.
We approximate the emergent spectrum by an equal-weight combination of the two polarization modes, while a full polarization-dependent transfer calculation is beyond the scope of this work.

\begin{figure}
\centerline{\includegraphics[width=0.5\textwidth]{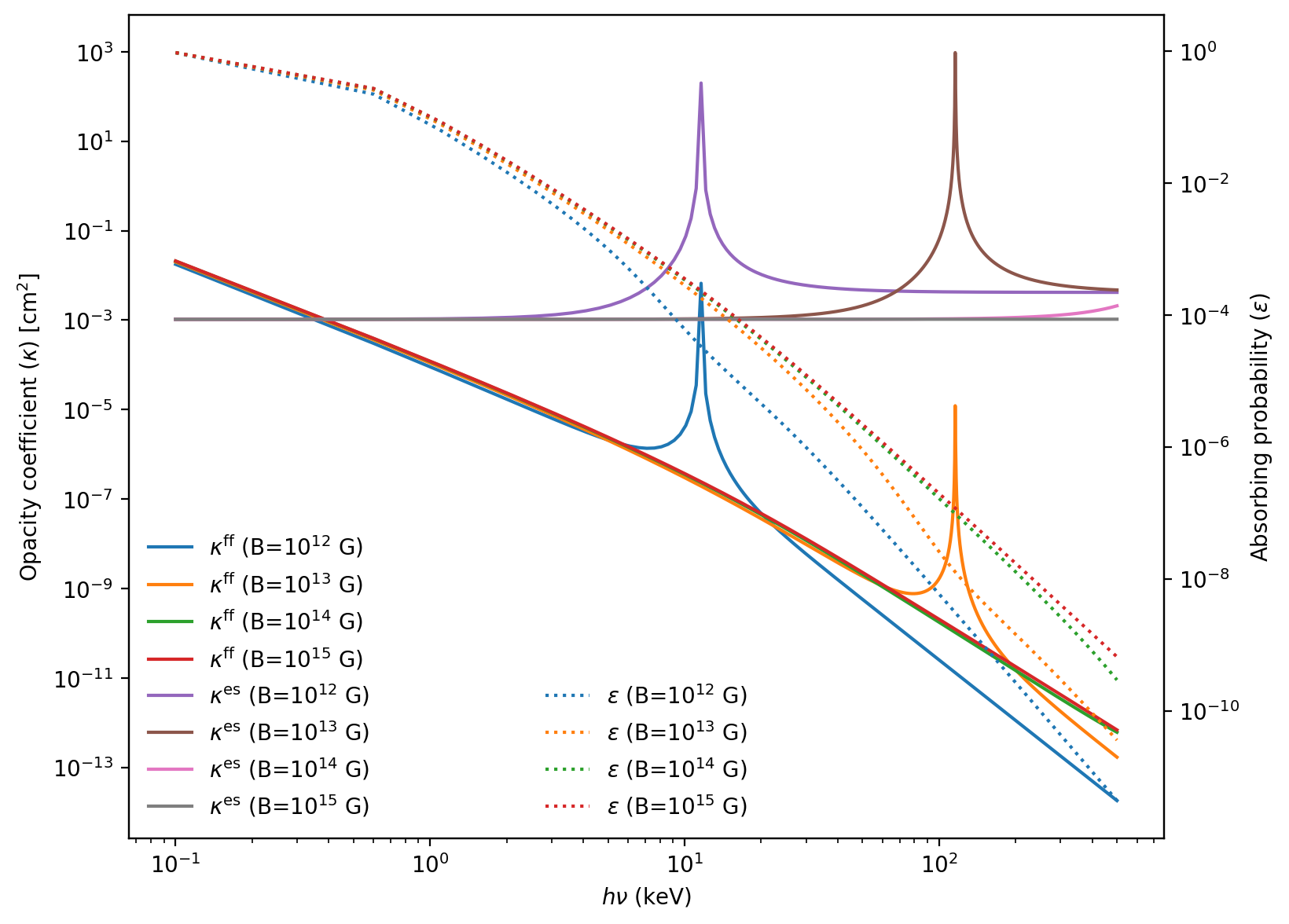}}
\caption{Opacity coefficients and absorbing probability as a function of photon energy $h\nu$ for different magnetic fields (B). The temperature $kT$=10 keV, Z=1, A=2, the angle between photon and magnetic field $\theta=45^\circ$, and the number density of electron $n_\mathrm{e}=3.14\times10^{21}~\rm cm^{-3}$.}
\label{fig:opacity}
\end{figure}

\begin{figure*}
\centerline{\includegraphics[width=\textwidth]{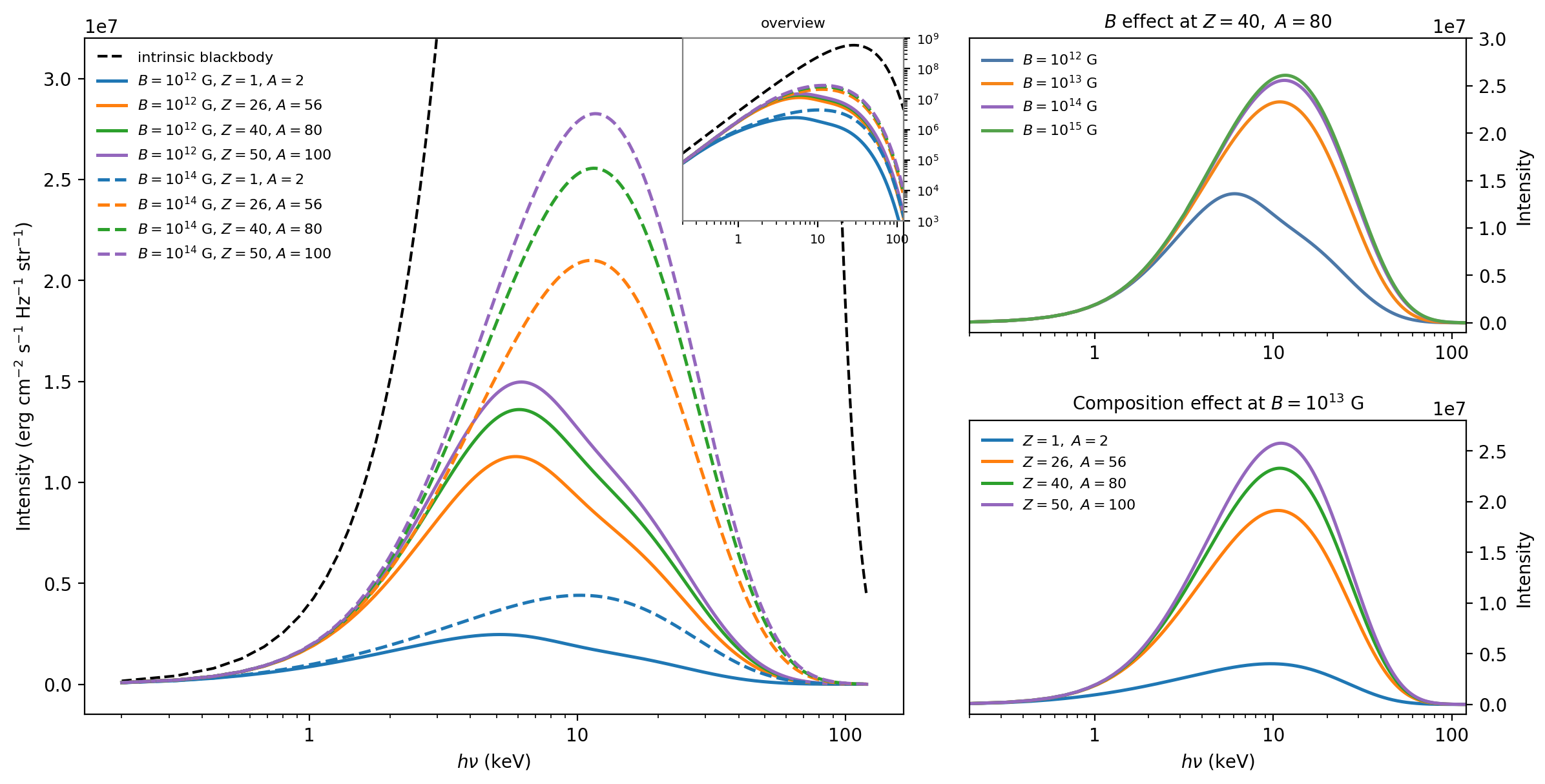}}
\caption{Intensity ($I_{\nu}^{\mathrm{E}}$) as a function of photon energy $h\nu$. 
The left panel shows representative spectra for different magnetic fields and ions, with the overview inset retaining the full intensity range including the intrinsic blackbody.
The upper-right panel shows the magnetic field effect for ions $(Z,A)=(40,80)$.
The lower-right panel shows the ionic effect for a magnetic field $B=10^{13}\,\rm G$.
The temperature is fixed at $kT=10$ keV, the angle between photon and magnetic field is $\theta=45^\circ$, and $n_\mathrm{e}=3.14\times10^{21}~\rm cm^{-3}$.}
\label{fig:intensity}
\end{figure*}

Figure \ref{fig:opacity} shows examples of scattering or absorption opacity as a function of photon energy for the magnetic field $B=10^{12}$--$10^{15}$ G, the temperature $kT=10$ keV, $Z=1$, $A=2$, the angle $\theta=45^\circ$, and the number density of electron estimated as \citep[e.g.,][]{Beloborodov2020ApJ,Kumar2020MNRAS},
\begin{equation}
\begin{aligned}
n_\mathrm{e}\approx&3.14\times10^{21}\:\mathrm{cm}^{-3}\left(\frac{\mathcal{M}}{10^8}\right)\left(\frac{P}{1\:\mathrm{s}}\right)^{-1}\\&\times\left(\frac{B_\mathrm{p}}{10^{15}\: \mathrm{G}}\right)\left(\frac{r}{R_\mathrm{NS}}\right)^{-3},
\end{aligned}
\end{equation}
where $P$ is the spin period of a magnetar, $B_\mathrm{p}$ is the surface dipolar magnetic field strength, $R_\mathrm{NS}$ is the radius of the NS, and $\mathcal{M}$ is the multiplicity \citep[e.g.,][]{MedinLai2010MNRAS,Beloborodov2013ApJ}.
At low energies ($\epsilon\sim1$), only the absorption effect is significant, corresponding to the blackbody (Rayleigh--Jeans) regime.
As energy increases, the spectrum transitions to a modified blackbody spectrum, as coherent scattering effects become increasingly important, corresponding to $y\lesssim1$.
Then, the spectrum as a function of photon energy for different magnetic fields/ions is shown in Figure \ref{fig:intensity}.
This indicates that the magnetic field strength affects the spectrum and may provide an approach to probing the magnetic field environment of the emitting region.
The charge number of nuclei also plays an important role in the spectrum. 
By contrast, the dependence on the ionic mass number is expected to be weaker, since neutrons do not participate directly in electromagnetic interactions and only indirectly affect the ionic cyclotron response.

\subsection{Inverse Compton effects}\label{subsec:inverse_compt}
When $y\gtrsim1$, inverse Compton will go to saturation and the spectrum, at high energies, becomes a Wien spectrum \citep{RybickiLightman1986book}:
\begin{equation}
\begin{aligned}
I_\nu^{\mathrm{W}}&=\frac{2h\nu^3}{c^2}e^{-\mu_\mathrm{c}}e^{-h\nu/kT}\\
&=n_\mathrm{c}\frac{2h\nu^3}{c^2}e^{-h\nu/kT},
\end{aligned}
\end{equation}
where $\mu_\mathrm{c}$ is the photon chemical potential, which parametrizes deviations from a Planck spectrum. 
Compton scattering redistributes photon energies while conserving the total number of photons and therefore cannot by itself drive the system to full thermodynamic equilibrium ($\mu_\mathrm{c}=0$). 
The normalization constant $n_\mathrm{c}=e^{-\mu_\mathrm{c}}$ is directly related to the total number of photons in the distribution \citep{Acharya2021MNRAS}.

\subsection{Observed flux spectrum}
The intensity considering absorption, coherent scattering, and inverse Compton effects may be estimated as \citep{RybickiLightman1986book}
\begin{equation}
\begin{aligned}
I_\nu&=I_{\nu}^{\mathrm{E}}\:+\:I_\nu^{\mathrm{W}} \\
&=I_\nu(\nu, \:kT, \:B, \:n_\mathrm{e}, \:Z, \:A, \:\mu_\mathrm{c}, \:\theta).
\end{aligned}
\end{equation}
Assuming the photons are isotropic before and after scattering, the differential number $\mathrm{d}N$ or the probability $\mathrm{d}N/N$ of photons in an interval ($\theta,\theta+\mathrm{d}\theta$) can be expressed as
\begin{equation}
\mathrm{d}N/N
=f(\theta)\mathrm{d}\theta
=\mathrm{d}\Omega/4\pi
=\sin\theta\,\mathrm{d}\theta/2,
\end{equation}
Therefore, we have
\begin{equation}
\begin{aligned}
I_\nu&=\int_{0}^{\pi}f(\theta) \cdot I_\nu(\theta)\,\mathrm{d}\theta \\
&=I_\nu(\nu, \:kT, \:B, \:n_\mathrm{e}, \:Z, \:A, \:\mu_\mathrm{c}).
\end{aligned}
\end{equation}

The observed flux $F_{\nu}$ [erg cm$^{-2}$ s$^{-1}$ Hz$^{-1}$] spectrum is then calculated by
\begin{equation}\label{eq:F_MEITP}
    F_\nu=\pi\left(\frac{l_\mathrm{p}}{D_\mathrm{L}}\right)^2\times I_\nu(\nu, \:kT, \:B, \:n_\mathrm{e}, \:Z, \:A, \:\mu_\mathrm{c}),
\end{equation}
where $l_\mathrm{p}$ is the length scale of the plasma and $D_\mathrm{L}$ is the luminosity distance from observer to the source.
In the following, we apply this MEITP spectrum model to the observed magnetar bursts.

\section{Data reduction and spectral fit}\label{sec:data_fit}
SGR~J1935+2154 is a Galactic magnetar that has attracted significant attention, having been observed for over 8 yr by multiple telescopes \citep[e.g.,][]{Israel2016MNRAS,Lin2020apj,Younes2020ApJ,Ridnaia2021NatAs,Borghese2022MNRAS,Cai2022b}, since its first discovery in 2014 \citep{Stamatikos2014gcn}, and in particular, has been found to host fast radio bursts \citep[e.g.][]{Bochenek2020Natur,CHIMEFRB2020nat,Li2021NatAs}.
To probe plasma properties that are sensitive to radiative transfer effects, such as the temperature, plasma density, magnetic field strength, and ionic charge, we focus on bright and relatively long-duration X-ray bursts.
These events provide sufficient photon statistics and optical depth for radiative transfer effects in a strongly magnetized plasma to become observationally relevant.
Accordingly, we select catastrophic energy release events, i.e., bursts that are brighter and longer than typical short bursts.
We obtain such bursts ($\gtrsim10^{41}~\rm erg$) from the history of Fermi/GBM and GECAM observations \citep[e.g.,][]{Lin2020apj,Lin2020apjl,Kaneko2021ApJ,Xie2022mnras,Rehan2023apj,Rehan2024ApJ,Xie2025ApJS,Rehan2025ApJS} and list them in Table \ref{tab:burst_list}.

We use the data reduction tool \textit{GBM Data Tools} \citep{GbmDataTools} for Fermi/GBM time tagged photon event (TTE) data and the \textit{GECAMTools} \citep{GECAMTools2024} for GECAM event-by-event (EVT) data to generate the time-integrated spectral data. The background spectra are inferred from data events in pre- and postburst time intervals (a few seconds) using a polynomial function to model the background.
The spectral duration and the detectors used are listed in Table \ref{tab:burst_list}.

The spectral data analysis tool \textit{elisa} is employed in this work \citep{ELISA}.
We perform a Bayesian fit using the MEITP model within the interested energy ranges (see Table \ref{tab:burst_list}).
The priors for the free parameters in Eq. \ref{eq:F_MEITP} are set as follows:
\begin{enumerate}
\item $kT$, the thermal equilibrium temperature of the plasma (e.g., fireball/bubble) in the observer frame. Uniform sampling in the range of 1--100 keV.
\item $n_\mathrm{e}$, the number density of electron in the plasma. Log-uniform sampling in the range of $10^{16}$--$10^{28}$ cm$^{-3}$.
\item $B$, the local magnetic field of the plasma. Log-uniform sampling in range of $10^{8}$--$10^{16}$ G.
\item $Z$, the ionic charge in the plasma. Uniform sampling in the range of 1--60. This range spans from hydrogen to the heaviest nuclei expected to be present in the outer crust of an isolated NS.
\item $\mu_\mathrm{c}$, the photon chemical potential. Uniform sampling in the range of 0--10.
\item $l_\mathrm{p}/D_\mathrm{L}$, the ratio of the plasma length scale to the luminosity distance. $l_\mathrm{p}$ is in units of km and $D_\mathrm{L}$ is in units of kpc. Log-uniform sampling in the range of $10^{-6}$--$10^{8}$.
\item $A$, the ionic mass number in the plasma.
We adopt a fixed relation $A=Z/0.5$ as a simplifying assumption.
Since neutrons do not directly contribute to radiative opacity in the considered energy range, variations in $A$ are expected to have a negligible impact on the spectral shape and are therefore not treated as an independent free parameter.
\end{enumerate}

The MEITP model fit results are listed in Table \ref{tab:specfit_result}. The spectrum and the posterior distributions of the parameters for each burst are plotted in Appendix \ref{apx_sec:spec_analysis}.

\begin{table*}
\caption{X-Ray bursts of SGR~J1935+2154 studied in This work.}
\label{tab:burst_list}
\begin{tabular}{lccccl}
\hline
Time & Instrument & Energy range & Duration & Detectors & Burst name \\
(UTC) & & (keV) & (s) & & \\
\hline
2016-06-26T13:54:30.722 & Fermi/GBM & 8-200 & 0.85 & n6,n9 & bn160626 \\
2020-04-27T18:32:41.595 & \nodata & \nodata & 2.19 & n6,n9 & bn200427a \\
2020-04-27T20:15:20.583 & \nodata & \nodata & 1.21 & na,n9,n2 & bn200427b \\
2021-09-10T00:45:46.875 & \nodata & \nodata & 1.39 & n8 & bn210910 \\
2021-09-11T15:06:43.188 & \nodata & \nodata & 0.49 & n6,n7,n9 & bn210911a \\
2021-09-11T15:15:25.373 & \nodata & \nodata & 1.17 & n9,n6 & bn210911b \\
2021-09-11T17:01:09.675 & \nodata & \nodata & 1.58 & n9,na & bn210911c \\
2021-12-24T03:42:34.341 & \nodata & \nodata & 1.30 & n5,n1,n3 & bn211224 \\
2022-01-12T08:39:25.279 & \nodata & \nodata & 1.04 & n1,n0,n2 & bn220112 \\
2021-01-30T17:40:54.750 & GECAM-B & 10-200 & 0.12 & g25H,g17H,g18H & bn210130 \\
2021-02-16T22:20:39.600 & \nodata & \nodata & 0.30 & g25H,g17H,g16H & bn210216 \\
\hline
\end{tabular}
\end{table*}

\begin{table*}
\caption{The fitting results of the MEITP model.}
\label{tab:specfit_result}
\begin{tabular}{lcccccccc}
\toprule
Burst name & $kT$ & $n_\mathrm{e}$ & $B$ & Z & $l_\mathrm{p}$ & $\mu_{\rm c}$ & $E_\mathrm{iso}$ & stat/dof \\
  & (keV) & ($10^{23}$cm$^{-3}$) & ($10^{12}$ G) & & (km) & & ($10^{40}$erg) & \\ 
\hline 
bn160626 & $15.45_{-0.69}^{+0.55}$ & $48.36_{-15.17}^{+41.61}$ & $2.31_{-0.09}^{+0.09}$ & $34.63_{-15.98}^{+15.89}$ & $10.77_{-0.21}^{+0.28}$ & $3.57_{-0.60}^{+0.61}$ & $19.26_{-0.10}^{+0.10}$ & 68.86/43 \\
bn200427a & $15.29_{-0.34}^{+0.33}$ & $54.86_{-16.34}^{+59.01}$ & $1.98_{-0.11}^{+0.09}$ & $34.73_{-17.93}^{+14.75}$ & $11.52_{-0.15}^{+0.16}$ & $3.41_{-0.30}^{+0.32}$ & $56.93_{-0.18}^{+0.18}$ & 281.56/44 \\
bn200427b & $14.14_{-0.54}^{+0.48}$ & $31.59_{-8.30}^{+19.37}$ & $1.43_{-0.14}^{+0.13}$ & $39.56_{-15.45}^{+12.96}$ & $10.69_{-0.16}^{+0.19}$ & $3.87_{-0.55}^{+0.55}$ & $17.21_{-0.07}^{+0.08}$ & 208.13/66 \\
bn210910 & $18.87_{-0.79}^{+0.81}$ & $4.43_{-0.94}^{+2.05}$ & $8.55_{-0.31}^{+0.34}$ & $43.71_{-13.88}^{+10.42}$ & $11.15_{-0.24}^{+0.23}$ & $3.87_{-0.38}^{+0.41}$ & $40.38_{-0.19}^{+0.19}$ & 54.56/21 \\
bn210911a & $15.80_{-1.14}^{+1.30}$ & $32.88_{-10.05}^{+25.44}$ & $1.05_{-0.14}^{+0.12}$ & $34.36_{-15.01}^{+13.91}$ & $9.43_{-0.33}^{+0.42}$ & $3.34_{-0.68}^{+0.94}$ & $7.34_{-0.04}^{+0.04}$ & 169.80/65 \\
bn210911b & $14.81_{-0.67}^{+0.52}$ & $50.44_{-14.50}^{+42.12}$ & $2.38_{-0.09}^{+0.09}$ & $36.49_{-16.65}^{+14.88}$ & $8.64_{-0.17}^{+0.23}$ & $3.51_{-0.63}^{+0.65}$ & $15.76_{-0.09}^{+0.09}$ & 99.86/42 \\
bn210911c & $14.32_{-0.87}^{+0.64}$ & $43.23_{-12.78}^{+36.96}$ & $1.47_{-0.19}^{+0.18}$ & $34.59_{-16.15}^{+13.44}$ & $8.15_{-0.16}^{+0.24}$ & $3.71_{-0.81}^{+0.76}$ & $14.66_{-0.08}^{+0.08}$ & 104.12/42 \\
bn211224 & $12.27_{-0.24}^{+0.30}$ & $13.33_{-3.22}^{+5.96}$ & $1.22_{-0.07}^{+0.08}$ & $40.96_{-12.18}^{+11.05}$ & $14.67_{-0.43}^{+0.44}$ & $2.17_{-0.13}^{+0.17}$ & $24.41_{-0.09}^{+0.08}$ & 154.90/67 \\
bn220112 & $14.90_{-1.10}^{+0.66}$ & $11.93_{-3.68}^{+11.33}$ & $1.57_{-0.12}^{+0.10}$ & $35.20_{-16.71}^{+16.45}$ & $8.94_{-0.29}^{+0.41}$ & $4.14_{-0.87}^{+0.77}$ & $8.14_{-0.05}^{+0.05}$ & 102.77/64 \\
bn210130 & $14.81_{-0.89}^{+1.34}$ & $5.41_{-2.74}^{+7.39}$ & $5.71_{-1.16}^{+1.83}$ & $30.65_{-16.10}^{+16.27}$ & $8.44_{-0.46}^{+0.49}$ & $2.51_{-0.42}^{+0.69}$ & $1.17_{-0.02}^{+0.02}$ & 69.53/70 \\
bn210216 & $10.08_{-0.86}^{+1.04}$ & $37.70_{-22.98}^{+56.72}$ & $2.95_{-0.61}^{+1.79}$ & $32.98_{-19.31}^{+17.36}$ & $8.41_{-0.42}^{+0.38}$ & $2.13_{-0.95}^{+1.60}$ & $1.41_{-0.02}^{+0.02}$ & 123.12/63 \\
\botrule
\end{tabular}
\tablecomments{The $l_\mathrm{p}$ and $E_\mathrm{iso}$ are inferred assuming the luminosity distance $D_\mathrm{L}$=9 kpc \citep{Zhong2020ApJ}. The isotropic energy $E_\mathrm{iso}$ is estimated within a specific energy range (see Table \ref{tab:burst_list}).}
\end{table*}

\section{Discussion} \label{sec:discussion}

\subsection{Spectral constraints on plasma composition and magnetic field}
\begin{figure}
\centerline{\includegraphics[width=0.5\textwidth]{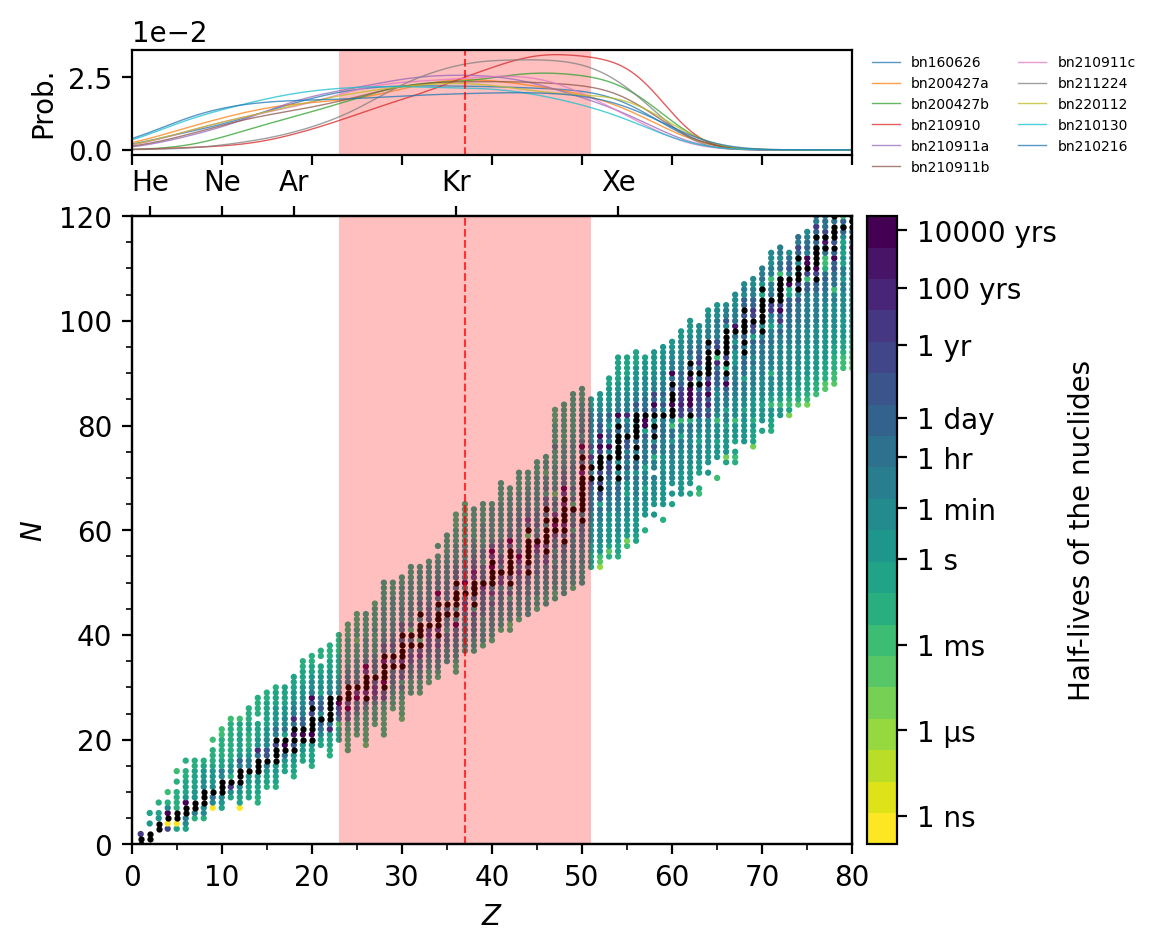}}
\caption{Valley of stability. Nuclide half-lives as a function of the number of protons $Z$, and neutrons $N$.
The upper panel shows the $Z$ posteriors from different bursts, with each burst represented by a distinctly colored line.
The vertical red dashed line and the shaded band represent the median value and the $1\sigma$ confidence level, respectively.
}
\label{fig:halflives}
\end{figure}

Our spectral analysis provides direct, model-dependent constraints on the physical conditions of the emitting plasma in bright magnetar bursts. 
Within the MEITP framework, the data strongly disfavor spectra produced by plasmas containing only light ions (e.g., H$^+$ and He$^{2+}$). 
Instead, the spectra are better reproduced when the radiative transfer is evaluated for an electron--ion plasma with larger effective ionic charge numbers, yielding an effective charge number of $Z = 37 \pm 14$, as shown in Figure~\ref{fig:halflives}. 
For this range of $Z$, the corresponding region of the nuclear chart contains many isotopes with half-lives longer than typical burst durations, indicating that radioactive decay is unlikely to significantly modify the plasma opacity and related radiative properties during the emission episode.
We emphasize that this constraint should be interpreted as an \emph{effective} charge number that characterizes the opacity in our model, rather than an identification of a unique nuclide.

In addition, the posterior distribution favors local magnetic fields of $B \sim 10^{12}$ G in the emission region. 
For a dipolar magnetic field, the magnetic strength as a function of radius $r$ can be estimated as $B(r)\sim B_\mathrm{p}~(r/R_\mathrm{NS})^{-3}$, where $B_\mathrm{p}=2.2\times10^{14}$ G is the surface dipolar magnetic field strength of SGR~J1935+2154 \citep{Israel2016MNRAS}.
This implies an emission radius $r \sim 6~R_\mathrm{NS}$, providing a physically interpretable link between the observed spectral shape and the location of the emitting plasma.

\subsection{Implications for crustal involvement and baryon loading}
The inferred preference for larger effective ionic charge numbers motivates considering scenarios in which ions contribute nonnegligibly to the radiative transfer in bright magnetar bursts.
In particular, this is naturally compatible with pictures in which crustal material participates in the formation of the emitting plasma, for instance, when the energy release occurs close to the surface and baryons are entrained into the fireball during its early evolution \citep[e.g.,][]{Thompson2001ApJ,vanPutten2016MNRAS,Ioka2020ApJ,Yang2021ApJ,Wada2023MNRAS,Cehula2024MNRAS,Patel2025ApJ}.
At the same time, the presence of heavy ions in the radiating plasma does not uniquely determine the triggering mechanism, since both crustal activity and magnetospheric processes may occur in the near-surface, strongly magnetized environment.

To assess whether such crustal involvement is energetically plausible, we compare the baryon content required by the radiative transfer fits with order-of-magnitude estimates of the material that could be released in crustal yielding events.
The total amount of baryons inferred to participate in the radiative transfer within the fitted emission volume is small compared to the amount of crustal material that could, in principle, be released during an energetic crustal yielding event. 
For example, the energy budget associated with breaking large crystallites in the crust can be substantial.
Quantitatively, it can be estimated as \citep{Baiko2018MNRAS,Zhang2022ApJ}
\begin{equation}\label{eq:nucle_eng}
Q\approx1.2\times10^{40}\:\mathrm{erg}\:\rho_{11}^{4/3}\:Z_{40}^2\:A_{122}^{-4/3}\:l_\mathrm{c}^3,
\end{equation}
where $\rho_{11}$ is the mass density in units of $10^{11}$g cm$^{-3}$, and $Z_{40}=Z/40$, $A_{122}=A/122$ are the ionic charge and mass number, respectively. $l_\mathrm{c}$ is the length scale of the cracked crustal volume in units of km.
Only a fraction of the released baryons needs to be mixed into the radiating plasma for the ionic charge to leave measurable imprints on the emergent spectrum.
Any excess baryons may escape along open field lines or fall back onto the star, potentially contributing to subsequent activity \citep[e.g.,][]{Kaneko2021ApJ,Xie2024ApJ}.

To place the inferred charge range in the context of crust physics, one may adopt an illustrative relation $Z/A\sim0.3$ near the neutron drip threshold, as expected for nuclei in the outer crust under standard ground-state nuclear models.
Under this assumption, our effective $Z$ values correspond to neutron numbers of order $N\sim 86$ (with substantial uncertainty), broadly consistent with expectations for neutron-rich nuclei near neutron drip.
We emphasize that this mapping relies on assumptions about the equilibrium nuclear composition of the crust, which depends on the crustal equation of state and nuclear mass models, and is not directly constrained by the burst spectra themselves.

Overall, these considerations indicate that baryon loading from the crust is energetically and spectrally plausible for bright magnetar bursts.
However, the dynamical evolution that brings a baryon-loaded fireball from its formation near the magnetar surface to an emission region tens of kilometers away remains uncertain.

\subsection{Consistency with independent constraints from the eclipse geometry}
The inferred emission location, tens of kilometers from the magnetar, and the local magnetic field strength together provide a useful context for fireball scenarios. 
In the classical trapped fireball scenario, the fireball is generally considered to be near the stellar surface. Following the energy release, a hot, optically thick plasma rapidly forms and subsequently expands along the magnetic field lines \citep[e.g.,][]{Thompson1995MNRAS,Thompson1996ApJ,Duncan1992ApJ,Paczynski1992AcA}. 
Alternatively, a fireball may form and radiate at higher altitudes in the magnetosphere, for instance, if the effective emission region is displaced from the surface or if the observed radiation is produced after some propagation and expansion. 
In this scenario, however, it remains unclear how crustal material can be lifted and mixed into the radiating plasma at such altitudes.
In this sense, our inferred height is compatible with a magnetospheric emission region.

We note that the present MEITP modeling does not explicitly include cyclotron absorption. If absorption becomes important under sufficiently high local density or appropriate propagation conditions, then the observed spectrum would be influenced by both heavy-ion effects and absorption, and the inferred constraints on the ionic composition could accordingly be modified. Nevertheless, the field strength and distance scale inferred here are broadly consistent with those derived from the eclipsed burst scenario and from cyclotron resonance considerations in our previous work \citep{Xie2026ApJ}. Therefore, although the detailed spectral interpretations differ from one to another, both analyses point to a magnetic field of order $10^{12}$~G and an emission region located at altitudes of several tens of kilometers, supporting a high-magnetospheric fireball picture.



\begin{acknowledgments}
This work is supported by the National Key R\&D Program of China (2021YFA0718500) and the National Natural Science Foundation of China (grant Nos. 12393811, 12273042).
\end{acknowledgments}

\bibliography{main}{}
\bibliographystyle{aasjournalv7}

\newpage
\appendix
\restartappendixnumbering

\section{Absorption and scattering}\label{apx_sec:opacity}
For the MEITP around a magnetar, the plasma dielectric tensor, in the coordinate system $\hat{X}$$\hat{Y}$$\hat{Z}$ with $\hat{B}$ along $\hat{Z}$, is given by \citep[e.g.,][]{HoLai2003MNRAS,Harding2006RPPh},
\begin{equation}\label{eq:dielec}
    [\epsilon^{(p)}]_{\hat{Z}=\hat{B}}=\begin{bmatrix}\varepsilon&\mathrm{i}g&0\\-\mathrm{i}g&\varepsilon&0\\0&0&\eta\end{bmatrix},
\end{equation}
where
\begin{equation}\label{eq:dielec_comp}
\begin{aligned}
\varepsilon\pm g&=1-\frac{v_e(1+\mathrm{i}\gamma_{ri})+v_i(1+\mathrm{i}\gamma_{re})}{(1+\mathrm{i}\gamma_{re}\pm u_e^{1/2})(1+\mathrm{i}\gamma_{ri}\mp u_i^{1/2})+\mathrm{i}\gamma_{ei}},
\\\eta&\simeq1-\frac{v_e}{1+\mathrm{i}(\gamma_{ei}+\gamma_{re})}-\frac{v_i}{1+\mathrm{i}(\gamma_{ei}+\gamma_{ri})}.
\end{aligned}
\end{equation}
In Eqs. \ref{eq:dielec} and \ref{eq:dielec_comp}, the dimensionless quantities are defined as,
\begin{equation}
    u_e=\frac{\omega_\mathrm{ce}^2}{\omega^2},\quad u_i=\frac{\omega_{ci}^2}{\omega^2},\quad v_e=\frac{\omega_{pe}^2}{\omega^2},\quad v_i=\frac{\omega_{pi}^2}{\omega^2},
\end{equation}
where $\omega_\mathrm{ce}=eB/(m_\mathrm{e}c)$ is the electron cyclotron frequency, $\omega_\mathrm{ci}=ZeB/(m_\mathrm{i}c)$ is the ion cyclotron frequency, $\omega_{pe}=(4\pi n_\mathrm{e}e^2/m_\mathrm{e})^{1/2}$ is the electron plasma frequency and $\omega_{pi}=(4\pi n_\mathrm{i}Z^2e^2/m_\mathrm{i})^{1/2}$ is the ion plasma frequency.
Note that positrons are not considered in the above plasma. Therefore, the electron number density is determined by the condition of charge neutrality, giving that $n_\mathrm{e}=Zn_\mathrm{i}$.
The dimensionless damping rates $\gamma_{ei}=\nu_{ei}/\omega,\gamma_{re}=\nu_{re}/\omega$ and $\gamma_ri=\nu_{ri}/\omega$ are given by,
\begin{equation}
\begin{aligned}
&\gamma_{ei} =\frac{Z^2n_ie^4}{\hbar\omega^2}\biggl(\frac{2\pi}{m_ekT}\biggr)^{1/2}(1-e^{-\hbar\omega/kT}) g_\alpha^{\mathrm{ff}}, \\
&\gamma_{re} =\frac{2e^2\omega}{3m_ec^3}, \\
&\gamma_{ri} =\frac{Z^2m_e}{Am_p}\gamma_{re}, 
\end{aligned}
\end{equation}
where $g_\alpha^{\mathrm{ff}}$ is the Gaunt factor which is given by \citep[e.g.,][]{Nagel1980ApJ},
\begin{equation}
\begin{aligned}
g_\perp&=\int_{-\infty}^\infty dx\exp\left(-\frac{\hbar\omega}{kT}\sinh^2x\right)C_1\left(\frac{\omega}{\omega_B}e^{2x}\right)\\g_\parallel&=\int_{-\infty}^\infty dx\exp\left(-\frac{\hbar\omega}{kT}\sinh^2x\right)2\frac{\omega}{\omega_B}e^{2x}C_0\left(\frac{\omega}{\omega_B}e^{2x}\right)
\end{aligned}
\end{equation}
where $g_{\pm1}^{\mathrm{ff}}=g_\perp^{\mathrm{ff}}$ and $g_0^{\mathrm{ff}}=g_\parallel^{\mathrm{ff}}$, and $C_0$, $C_1$ are
Coulomb matrix elements \citep[e.g.,][]{Virtamo1975NCimB}.

In strong magnetic fields, vacuum polarization can significantly influence photon propagation.
Vacuum polarization contributes a correction to the dielectric tensor $\epsilon$ and the inverse magnetic permeability tensor ($\mu^{-1}\equiv\bar{\mu}$):
\begin{equation}
\begin{aligned}
\Delta\epsilon^{(v)}=\hat{a}\boldsymbol{I}+q\hat{\boldsymbol{B}}\hat{\boldsymbol{B}},\\
\Delta\bar{\mu}^{(v)}=\hat{a}\boldsymbol{I}+m\hat{\boldsymbol{B}}\boldsymbol{\hat{B}},
\end{aligned}
\end{equation}
where $\boldsymbol{I}$ is unit tensor, and $\hat{a}$, $q$ and $m$ are functions of the dimensionless magnetic field parameter $b\equiv B/B_\mathrm{Q}$.
The vacuum polarization coefficients are estimated by \citep{Harding2006RPPh},
\begin{equation}\label{eq:vacu_polar_coeff}
\begin{aligned}
\hat{a}&\approx-\frac{2\alpha_\mathrm{F}}{9\pi}\ln\left(1+\frac{b^2}{5} \frac{1+0.25487 b^{3/4}}{1+0.75 b^{5/4}}\right), \\
q&\approx\frac{7\alpha_\mathrm{F}}{45\pi} b^{2} \frac{1+1.2 b}{1+1.33 b+0.56 b^{2}}, \\
m&\approx-\frac{\alpha_\mathrm{F}}{3\pi} \frac{b^2}{3.75+2.7 b^{5/4}+b^2},
\end{aligned}
\end{equation}
where $\alpha_\mathrm{F}=e^2/\hbar c \simeq1/137$ is the fine-structure constant. These approximation calculations can span from weak field to high field ($b\gg1$).

Therefore, the ellipticity $K_j$ in Eq. \ref{eq:cyc_comp} can be calculated as follows \citep{HoLai2003MNRAS,Harding2006RPPh}:
\begin{equation}
    K_j=\beta_{\mathrm{p}}\left[1+(-1)^j\left(1+\frac r{\beta_{\mathrm{p}}^2}\right)^{1/2}\right],
\end{equation}
where $r=1+(m/a)\sin^2\theta$.
The polarization parameter $\beta_{\mathrm{p}}$ is given by,
\begin{equation}
    \beta_{\mathrm{p}}=-\frac{\varepsilon^{\prime2}-g^2-\varepsilon^{\prime}\eta^{\prime}\left(1+m/a\right)}{2g\eta^{\prime}}\frac{\sin^2\theta}{\cos\theta},
\end{equation}
where $\hat{a}=a-1$, $\varepsilon^{\prime}=\varepsilon+\hat{a}$, and $\eta^{\prime}=\eta+\hat{a}+q$.
In the transverse-mode approximation, we have $K_{z,j}=0$ and $A_\alpha=1$ \citep{HoLai2003MNRAS}.
Thus, one may estimate the opacities in equations \ref{eq:kap_scat} and \ref{eq:kap_abs}.

\section{Details of spectral analysis}\label{apx_sec:spec_analysis}
\begin{figure}
\centering
\includegraphics[width=0.45\textwidth]{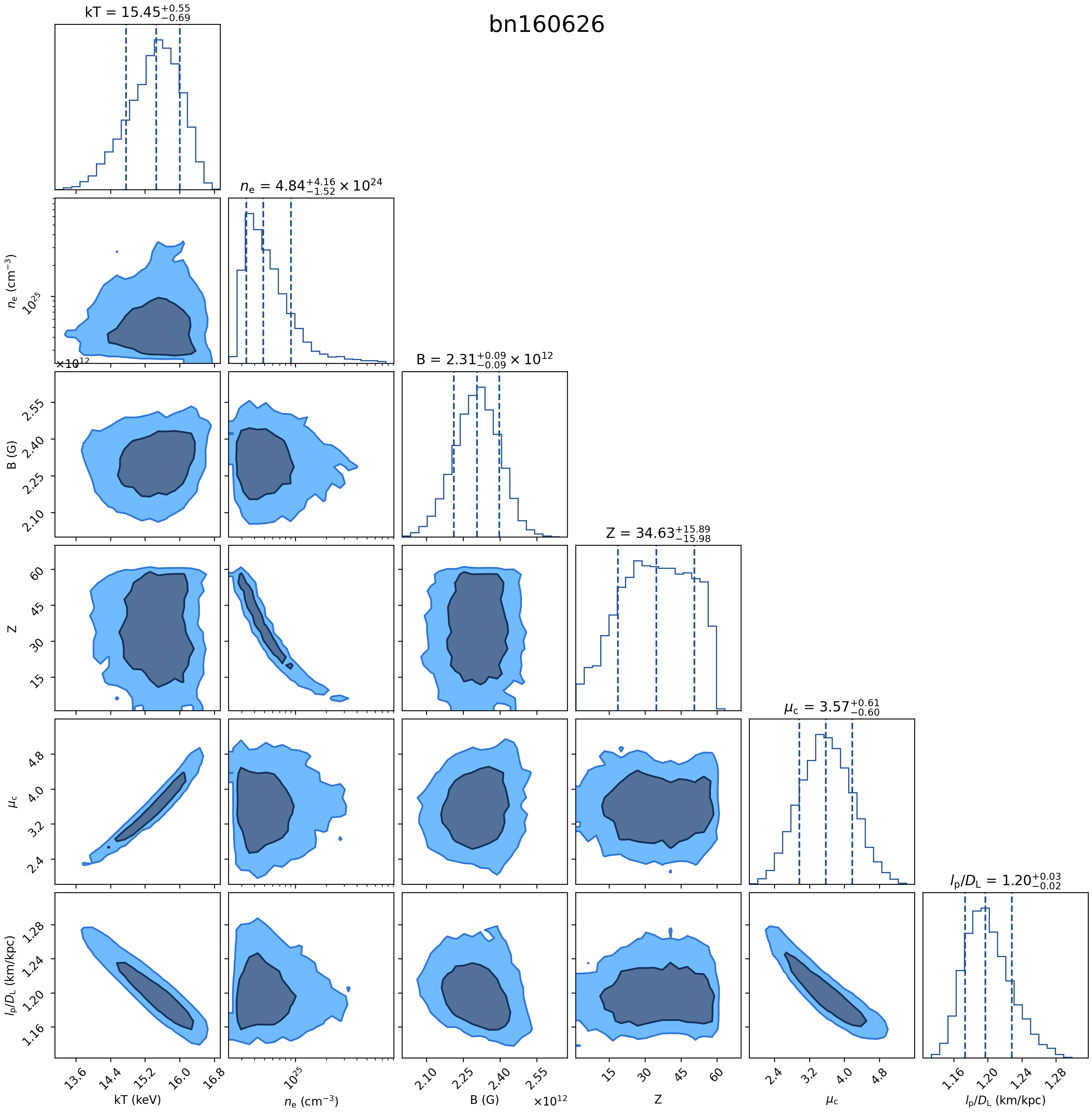}
\caption{Corner plots. The full set is available online.}
\label{fig:corner_set}
\end{figure}

\begin{figure}
\centering
\includegraphics[width=0.45\textwidth]{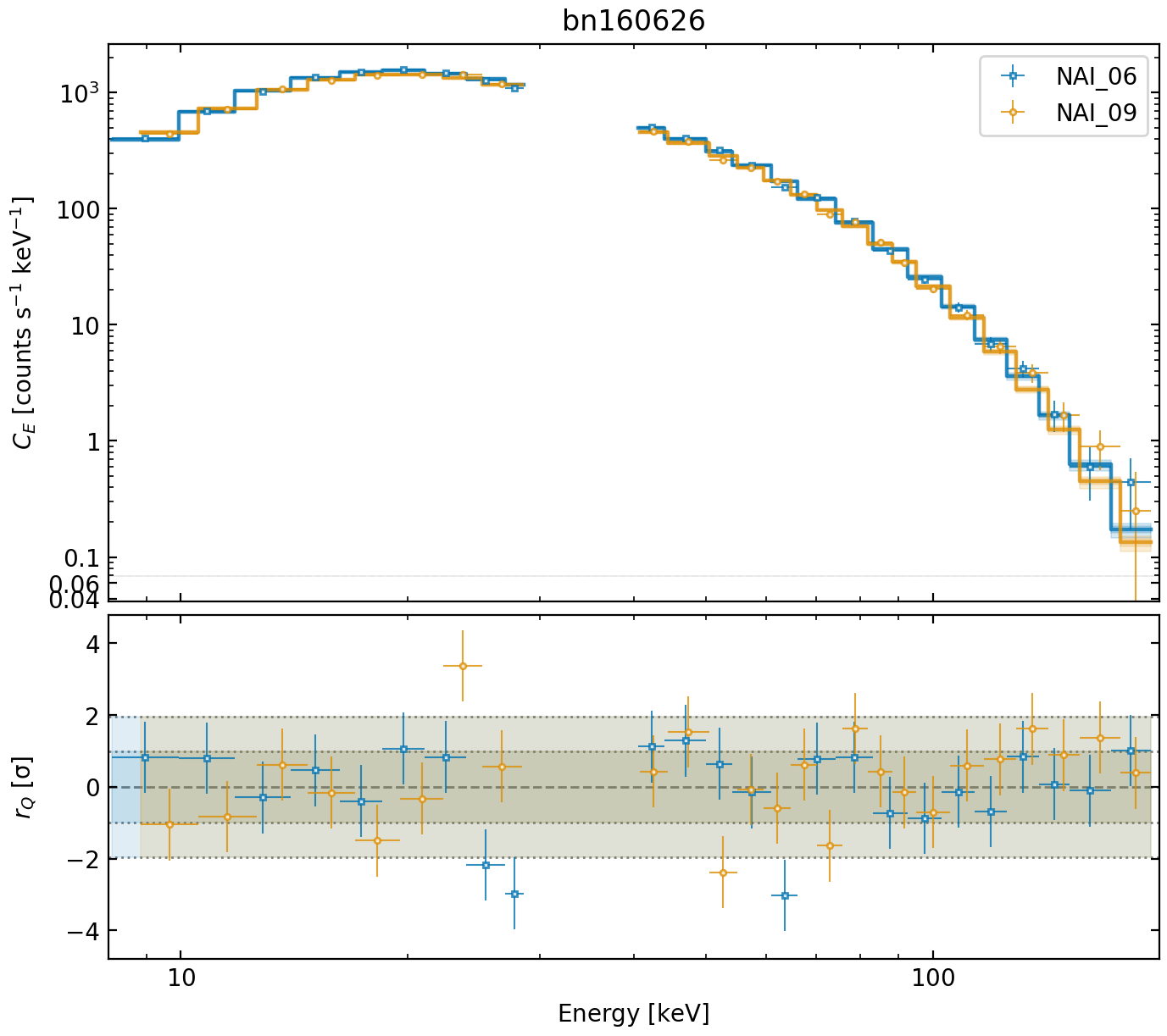}
\caption{Spectral fits. The full set is available online.}
\label{fig:spec_set}
\end{figure}

The spectrum and the posterior distributions of the parameters for each burst are plotted in Figs \ref{fig:corner_set}-\ref{fig:spec_set}.

\figsetstart
\figsetnum{B1}
\figsettitle{Corner plots for individual bursts}

\figsetgrpstart
\figsetgrpnum{1}
\figsetgrptitle{bn160626}
\figsetplot{figure/Corner_bn20160626_135430.png}
\figsetgrpend

\figsetgrpstart
\figsetgrpnum{2}
\figsetgrptitle{bn200427a}
\figsetplot{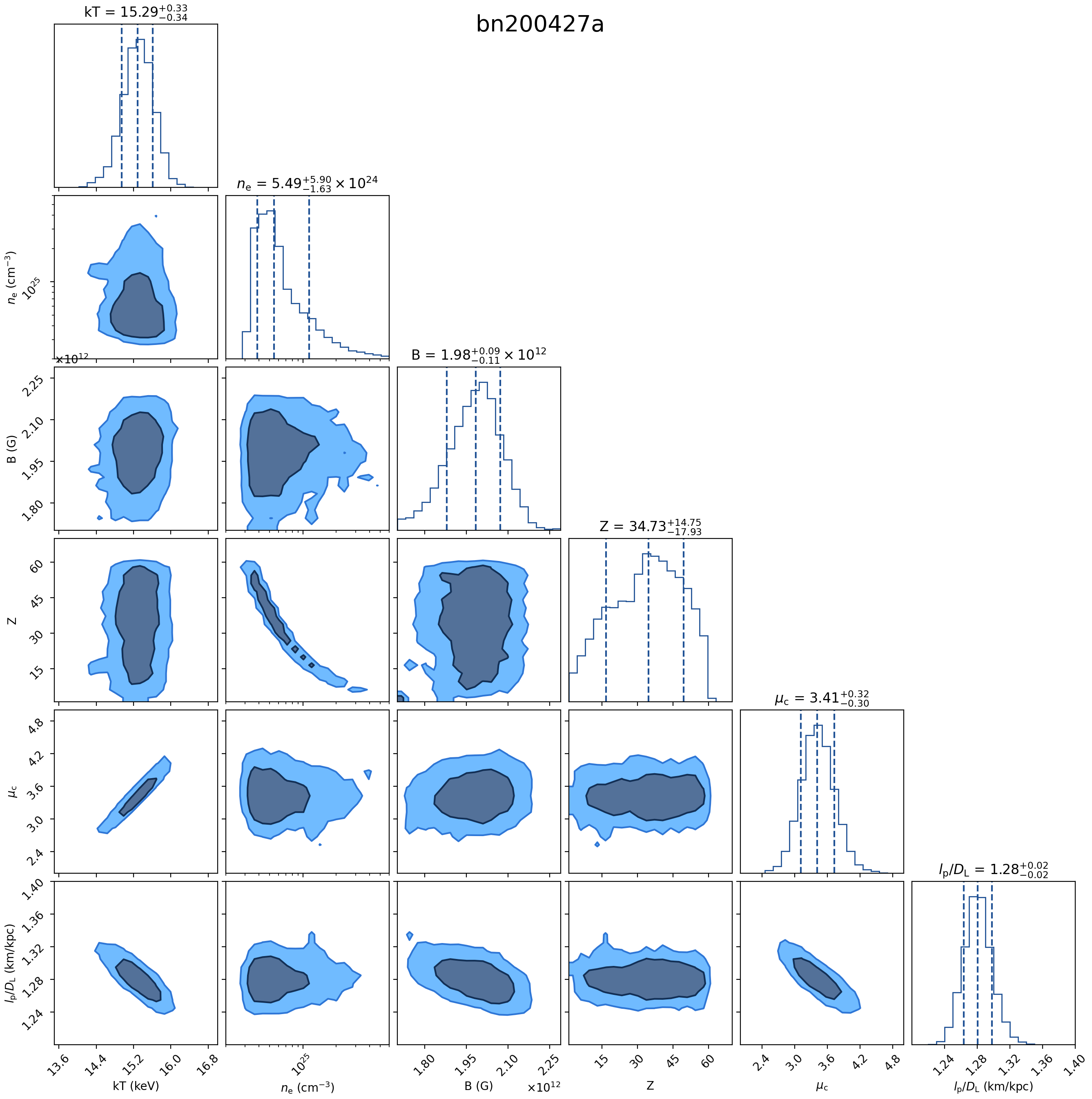}
\figsetgrpend

\figsetgrpstart
\figsetgrpnum{3}
\figsetgrptitle{bn200427b}
\figsetplot{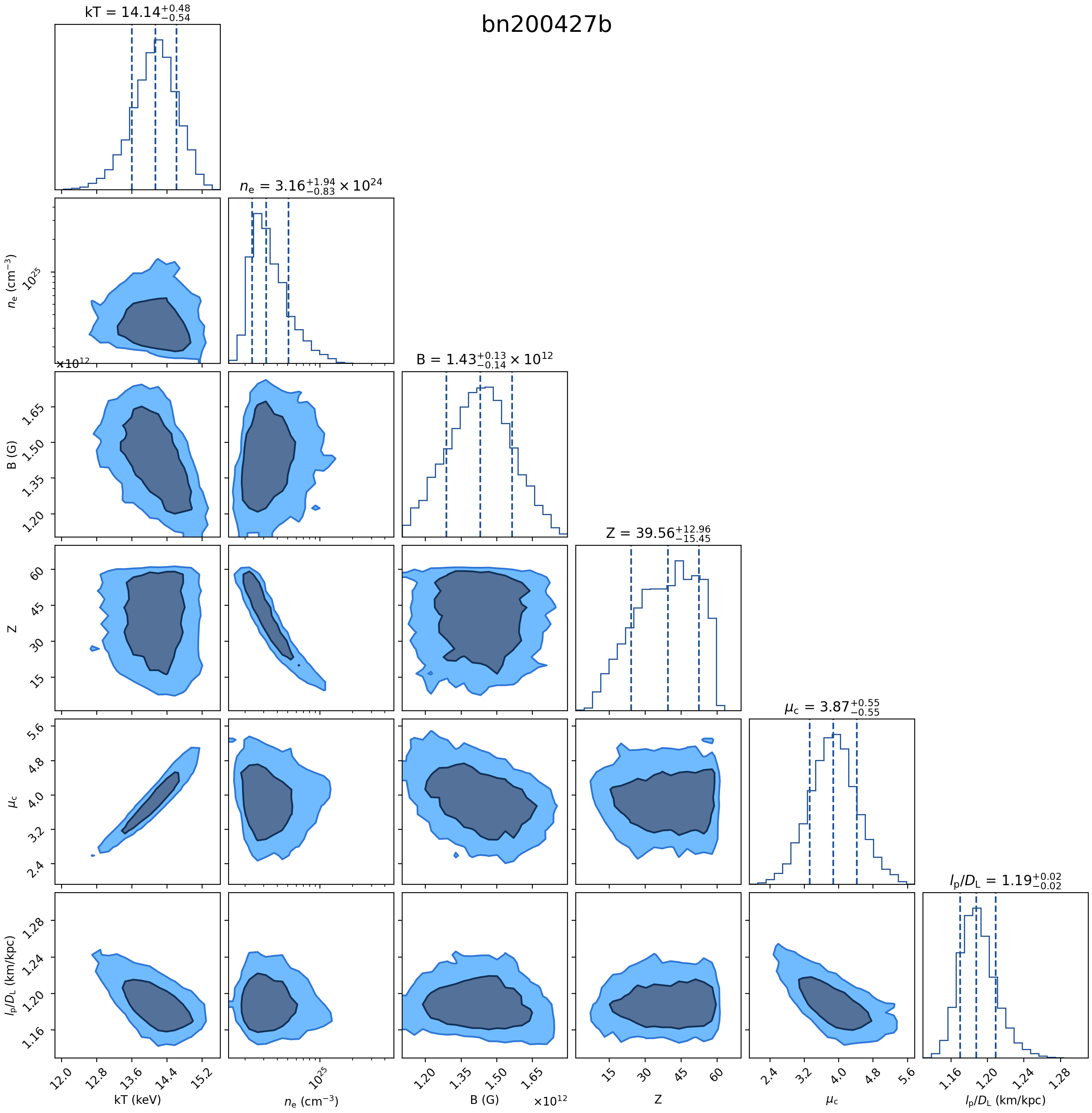}
\figsetgrpend

\figsetgrpstart
\figsetgrpnum{4}
\figsetgrptitle{bn210910}
\figsetplot{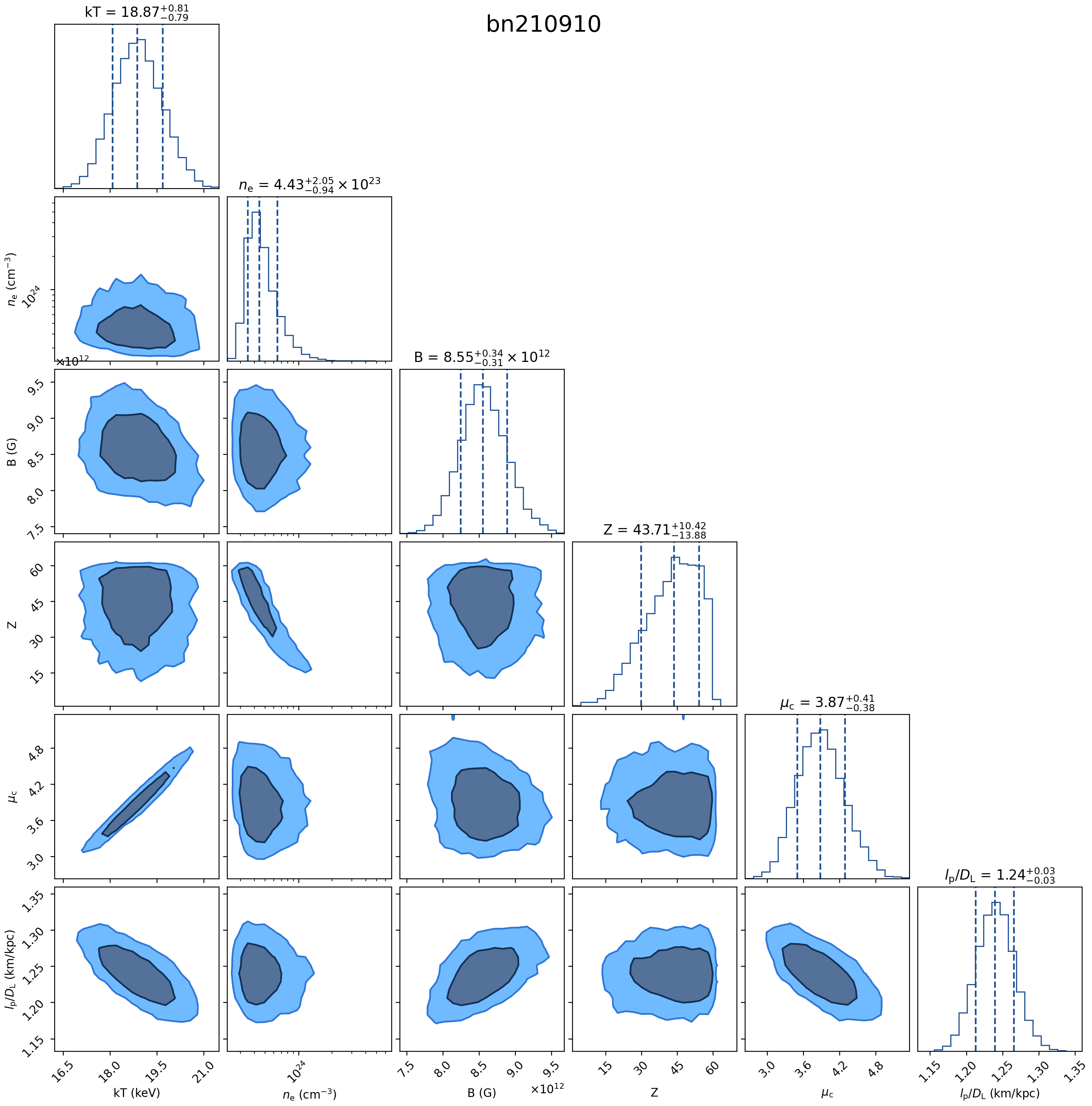}
\figsetgrpend

\figsetgrpstart
\figsetgrpnum{5}
\figsetgrptitle{bn210911a}
\figsetplot{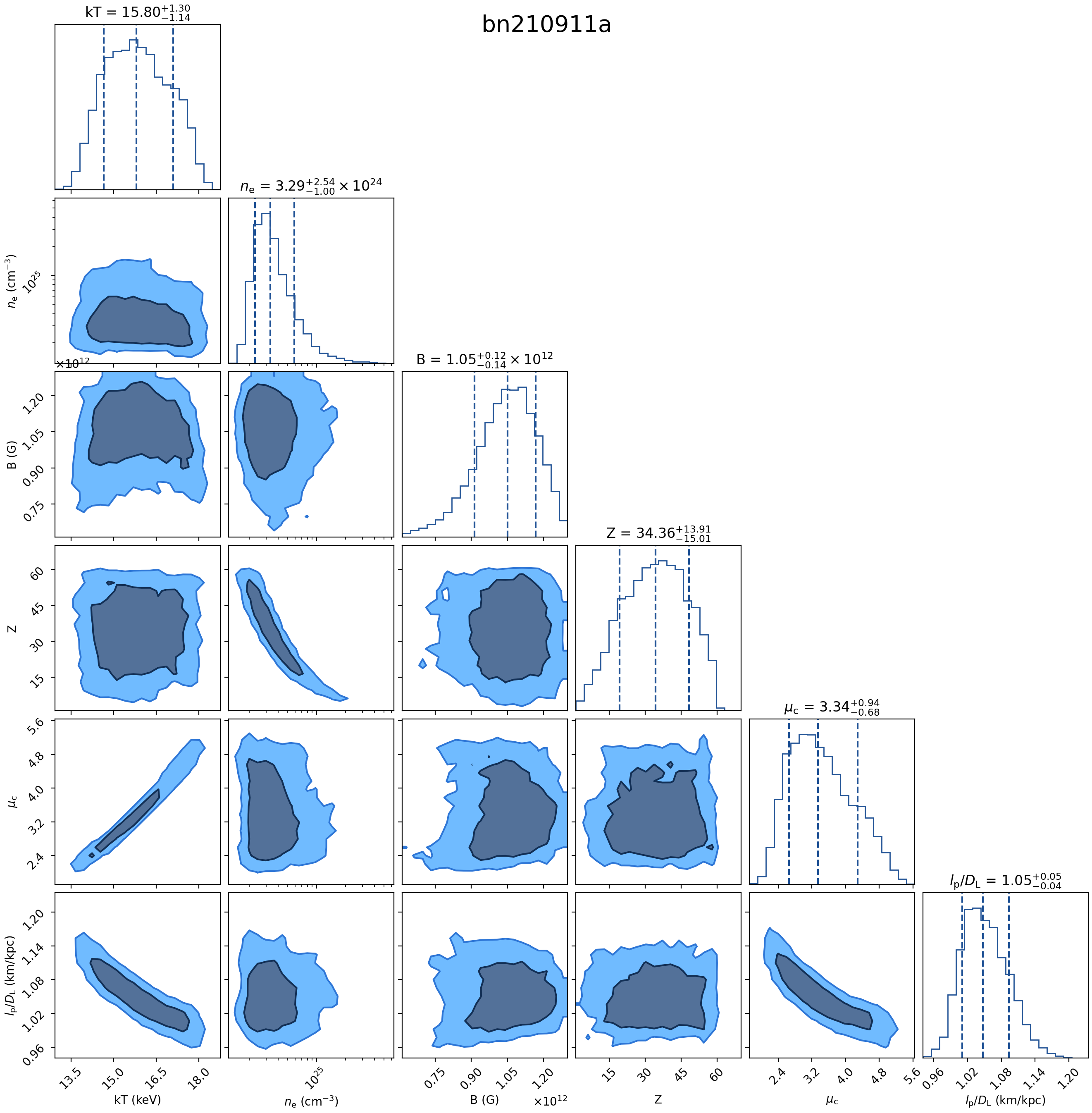}
\figsetgrpend

\figsetgrpstart
\figsetgrpnum{6}
\figsetgrptitle{bn210911b}
\figsetplot{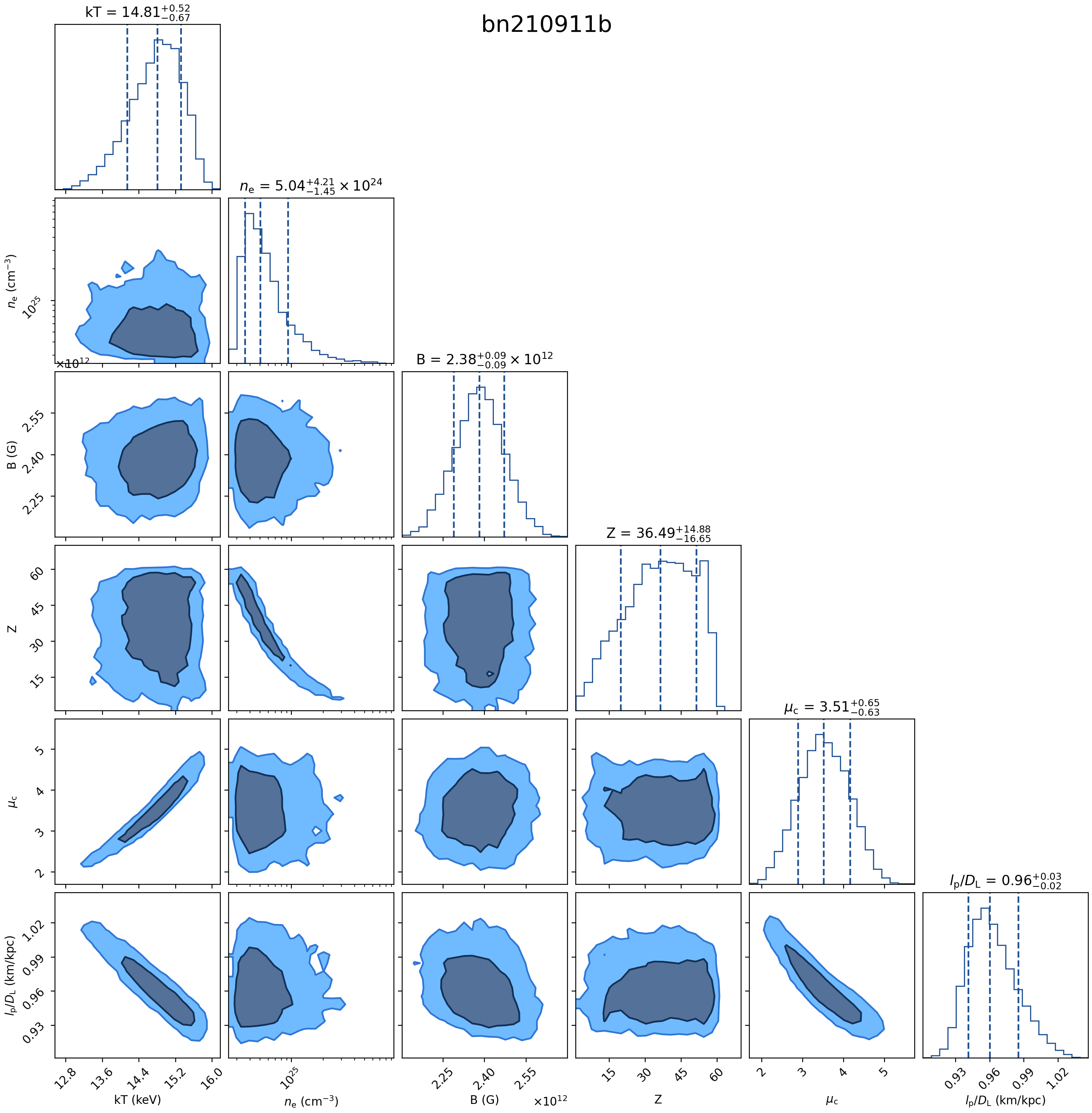}
\figsetgrpend

\figsetgrpstart
\figsetgrpnum{7}
\figsetgrptitle{bn210911c}
\figsetplot{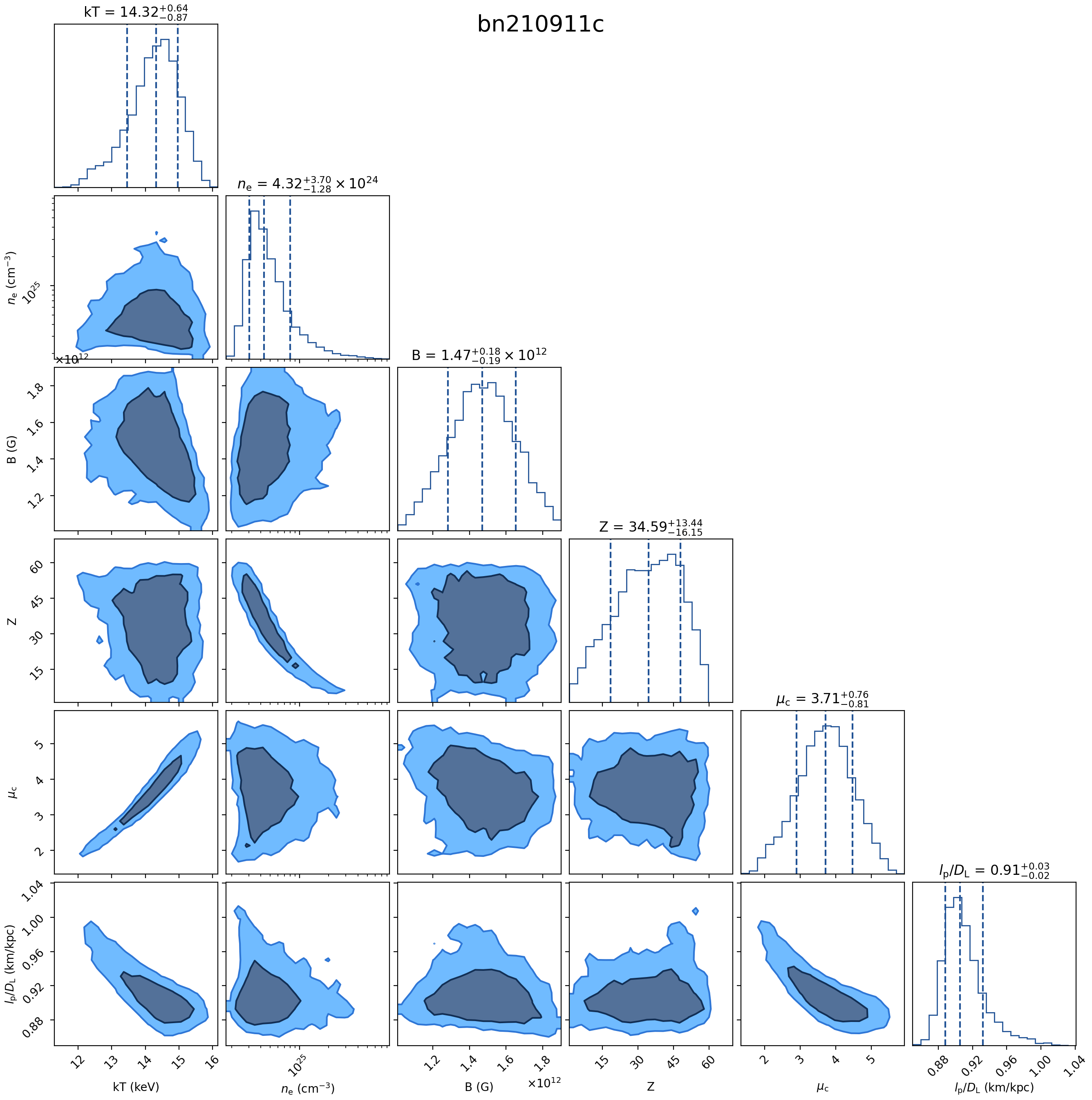}
\figsetgrpend

\figsetgrpstart
\figsetgrpnum{8}
\figsetgrptitle{bn211224}
\figsetplot{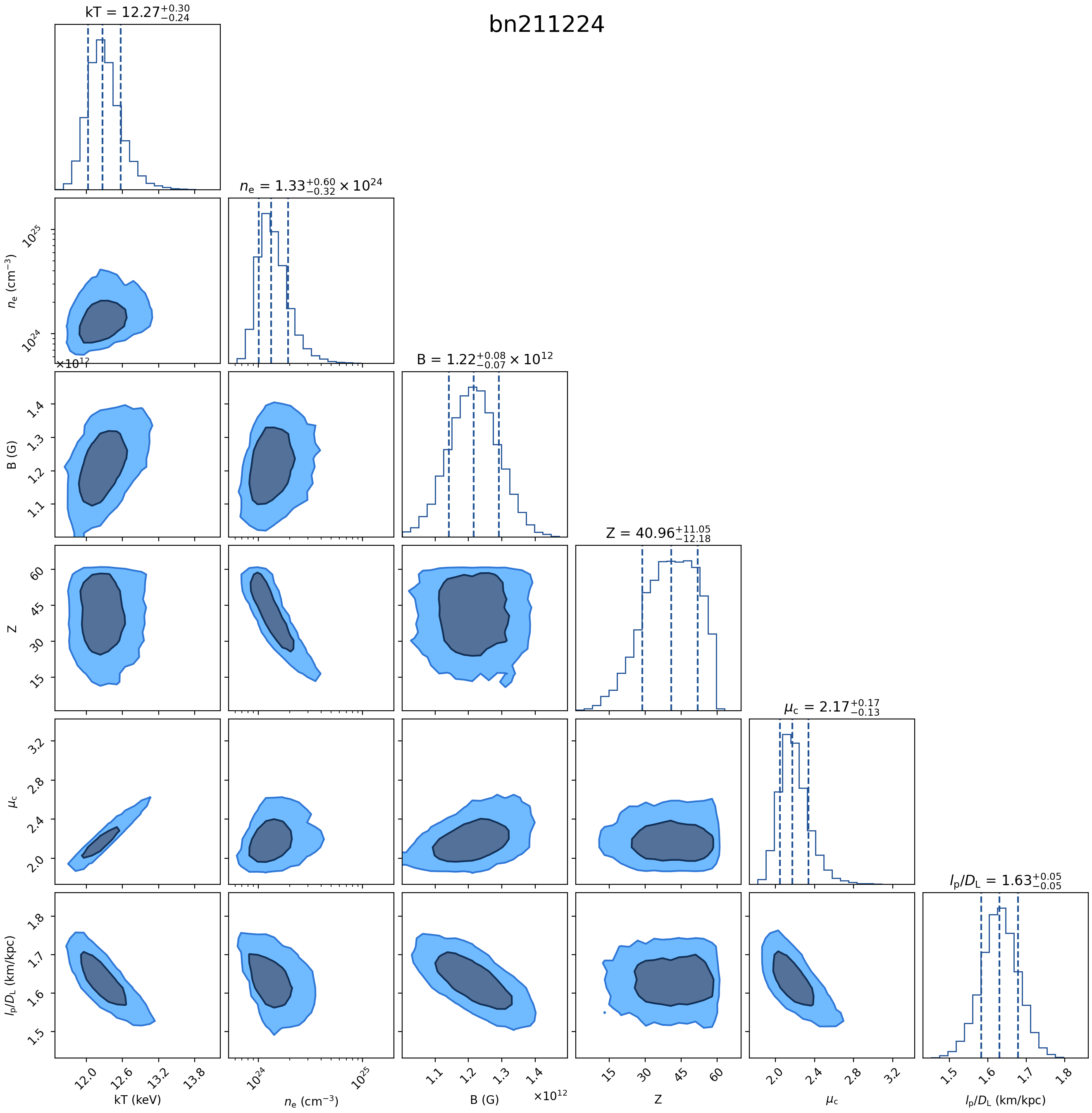}
\figsetgrpend

\figsetgrpstart
\figsetgrpnum{9}
\figsetgrptitle{bn220112}
\figsetplot{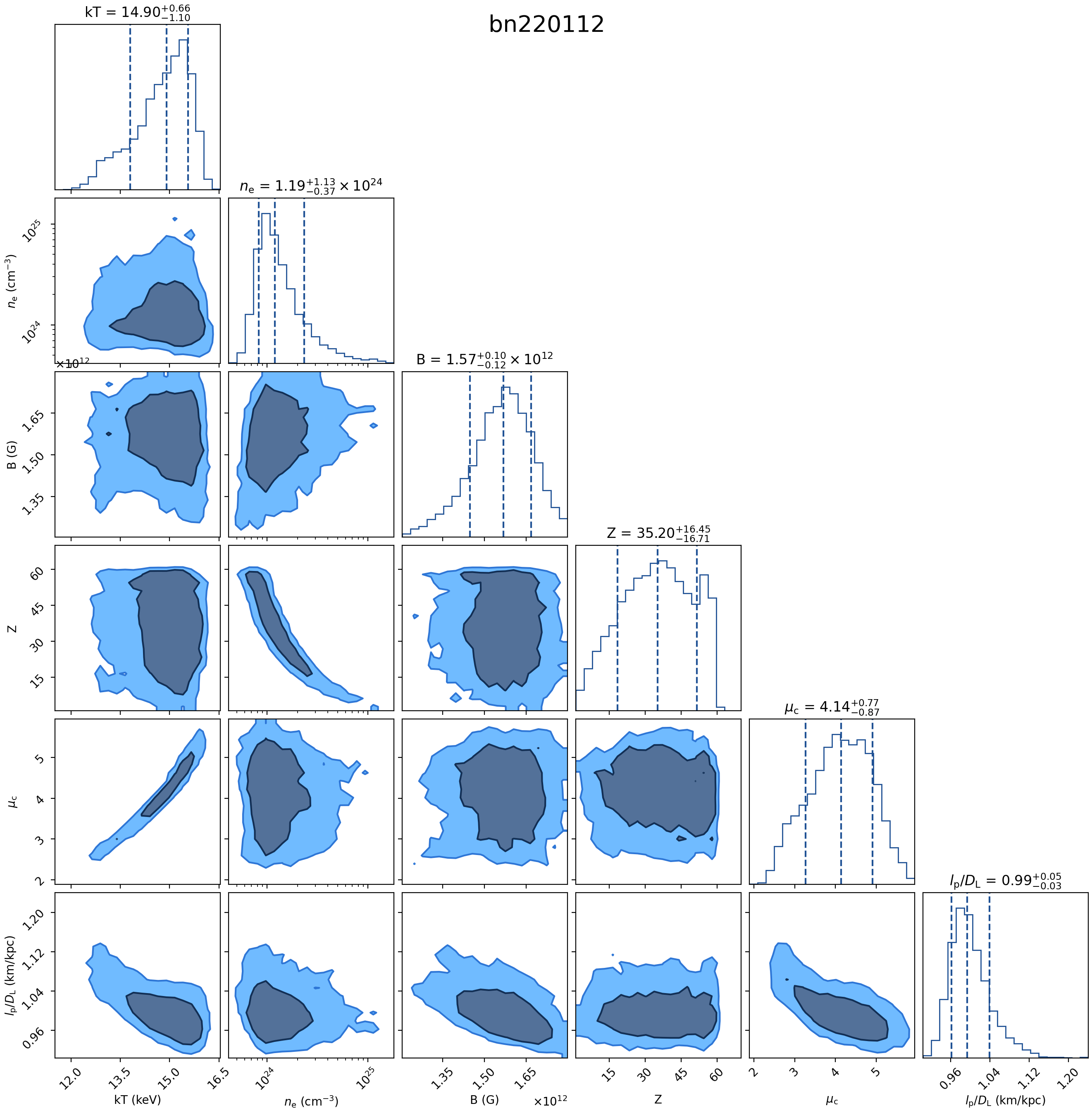}
\figsetgrpend

\figsetgrpstart
\figsetgrpnum{10}
\figsetgrptitle{bn210130}
\figsetplot{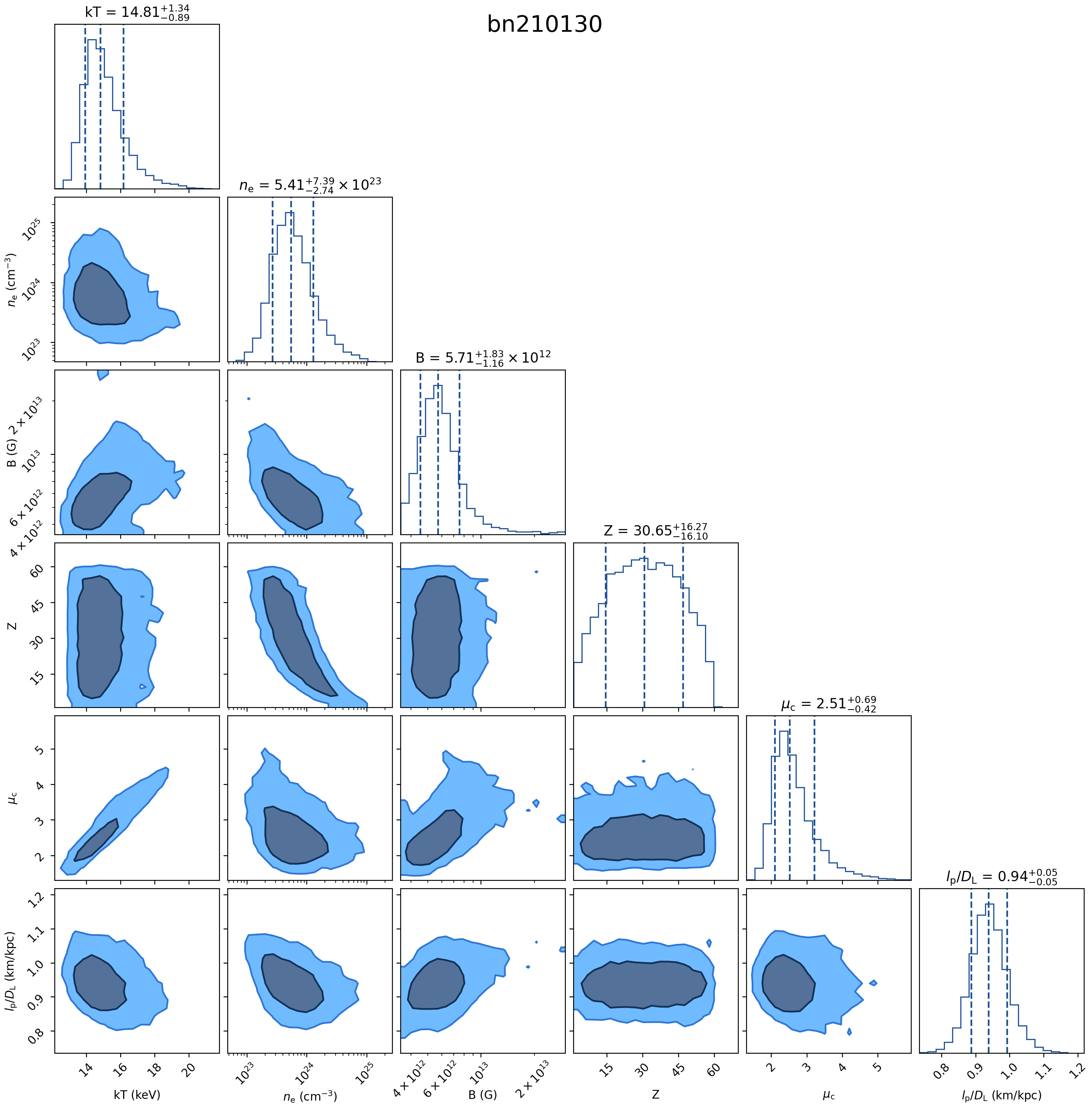}
\figsetgrpend

\figsetgrpstart
\figsetgrpnum{11}
\figsetgrptitle{bn210216}
\figsetplot{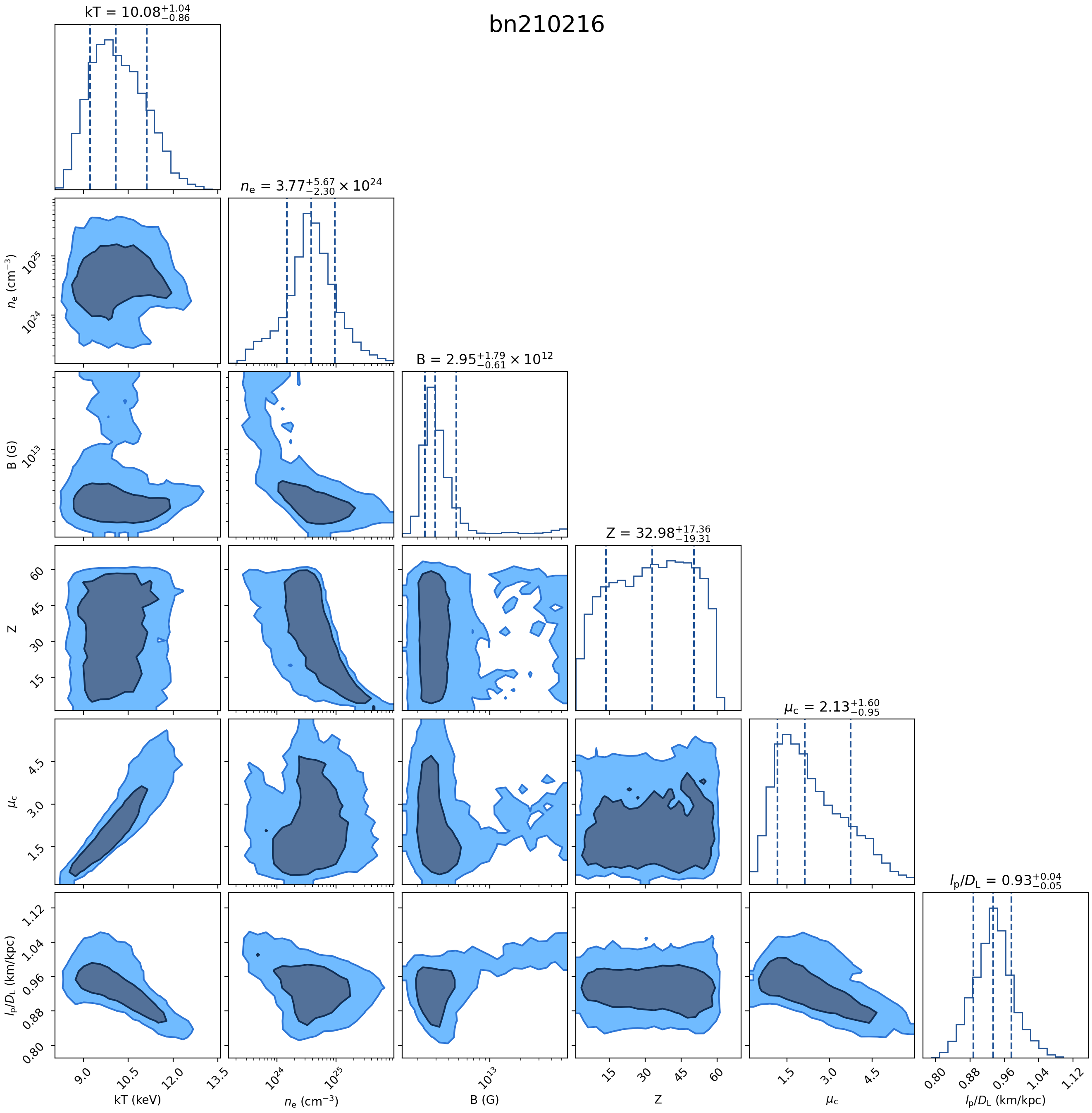}
\figsetgrpend

\figsetend

\figsetstart
\figsetnum{B2}
\figsettitle{Spectral fits for individual bursts}

\figsetgrpstart
\figsetgrpnum{1}
\figsetgrptitle{bn160626}
\figsetplot{figure/Spec_bn20160626_135430.png}
\figsetgrpend

\figsetgrpstart
\figsetgrpnum{2}
\figsetgrptitle{bn200427a}
\figsetplot{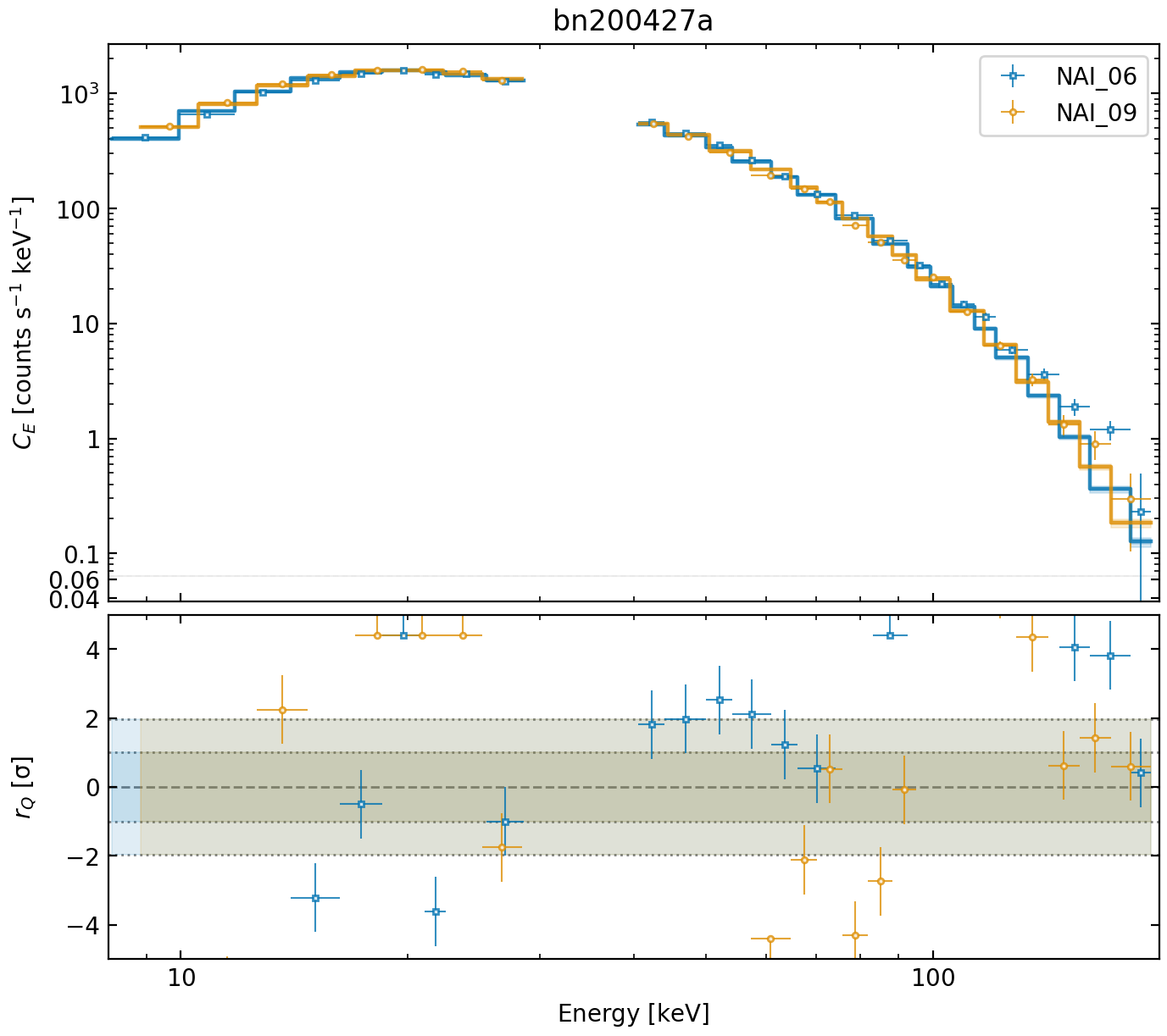}
\figsetgrpend

\figsetgrpstart
\figsetgrpnum{3}
\figsetgrptitle{bn200427b}
\figsetplot{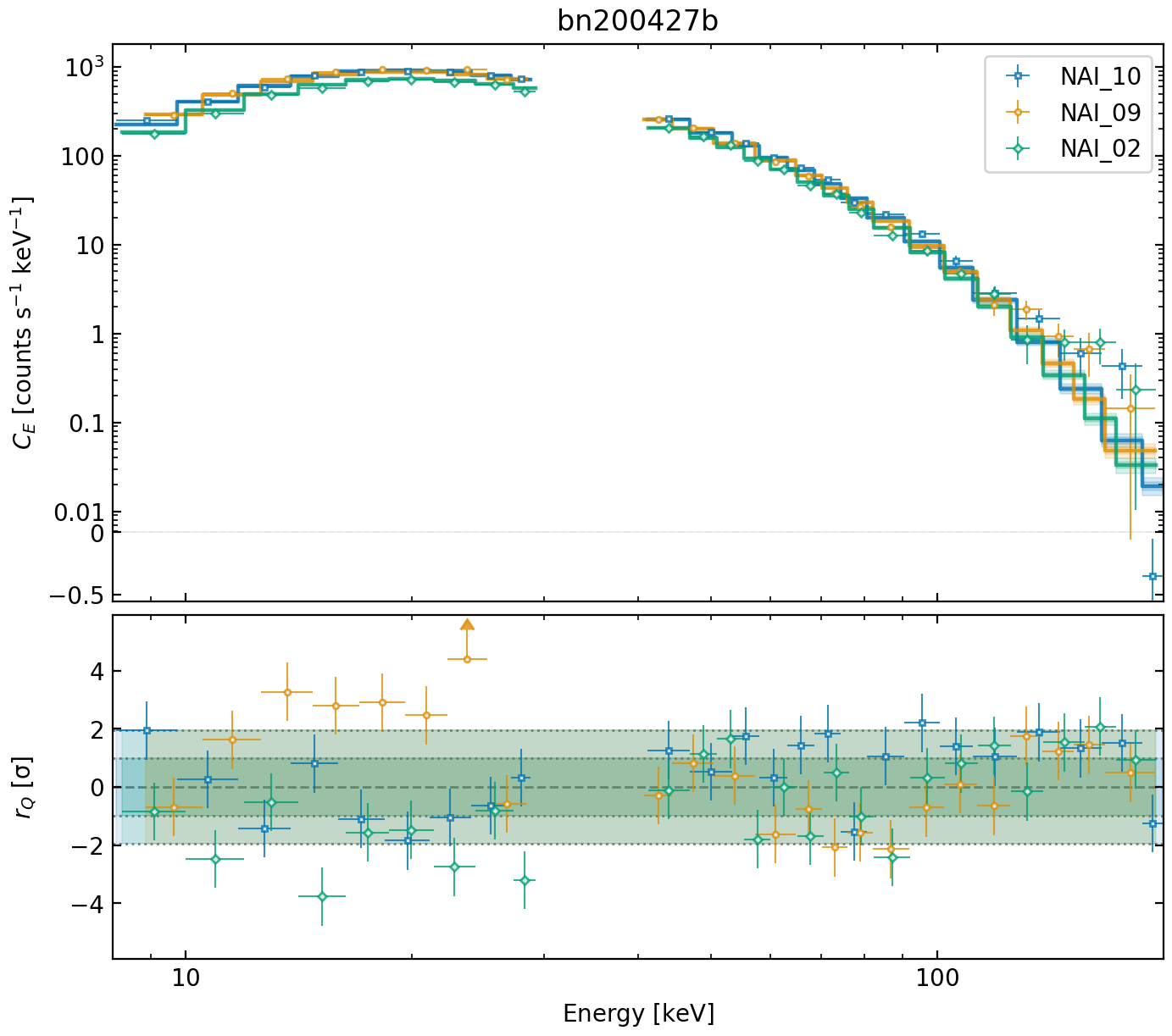}
\figsetgrpend

\figsetgrpstart
\figsetgrpnum{4}
\figsetgrptitle{bn210910}
\figsetplot{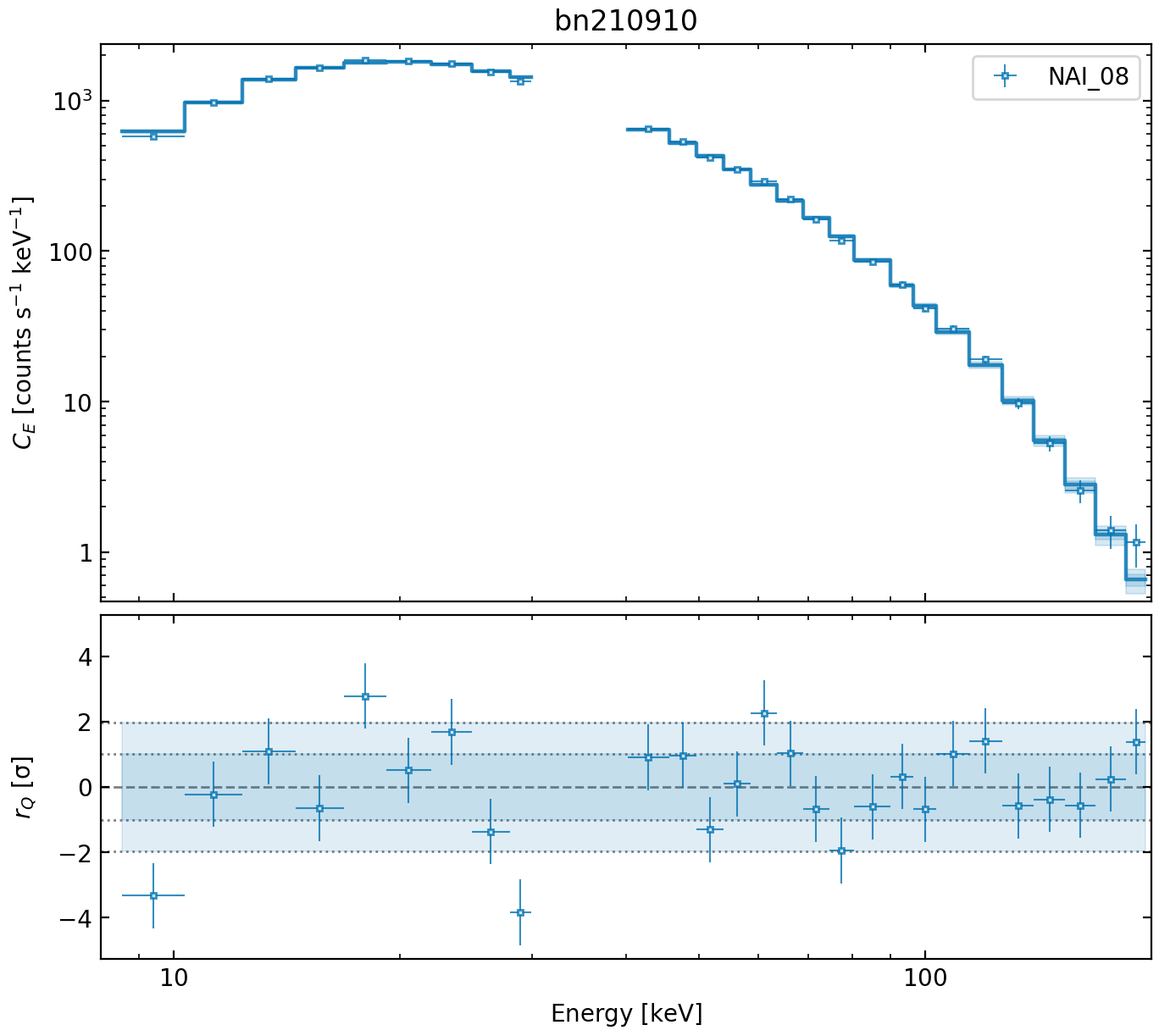}
\figsetgrpend

\figsetgrpstart
\figsetgrpnum{5}
\figsetgrptitle{bn210911a}
\figsetplot{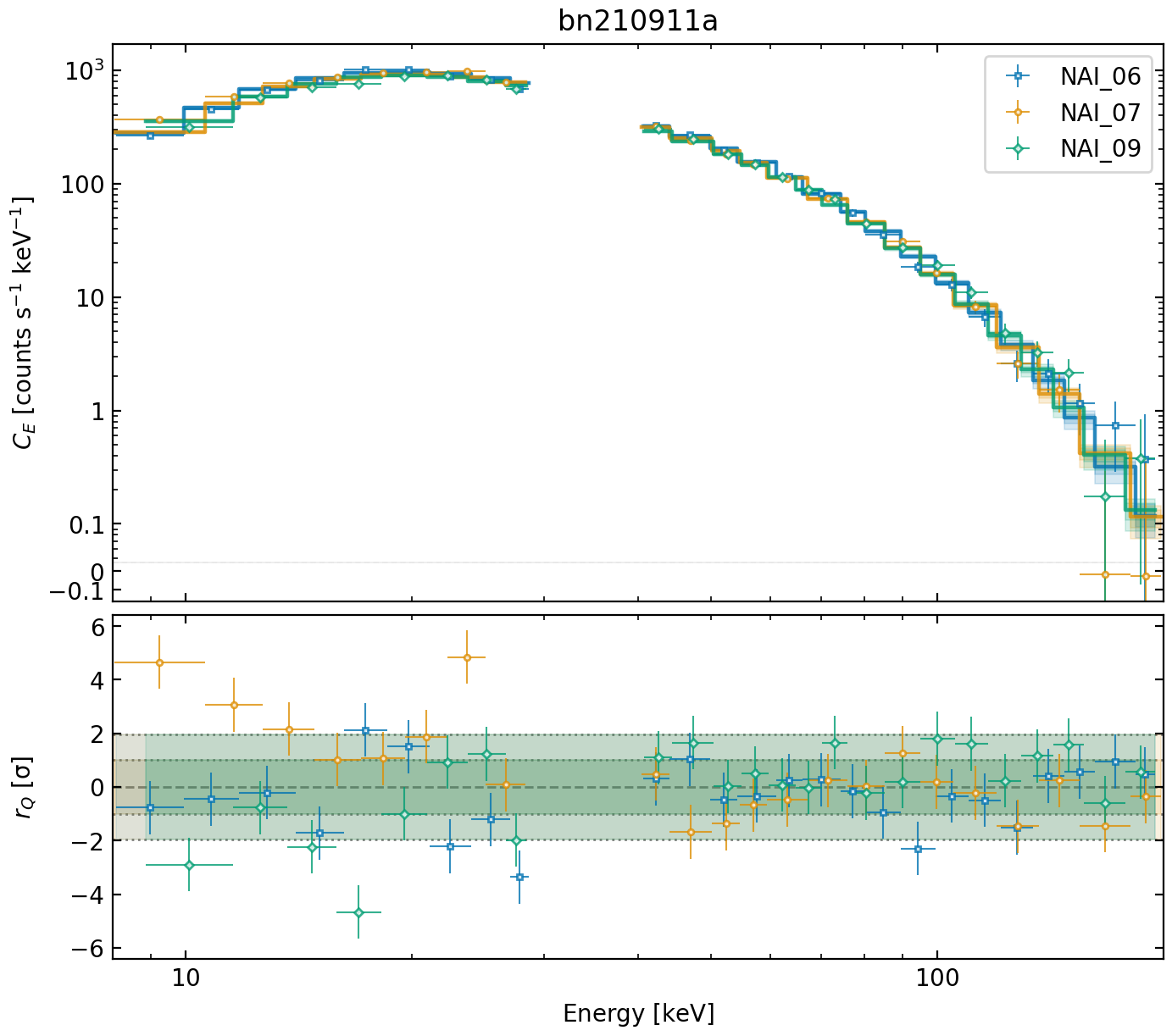}
\figsetgrpend

\figsetgrpstart
\figsetgrpnum{6}
\figsetgrptitle{bn210911b}
\figsetplot{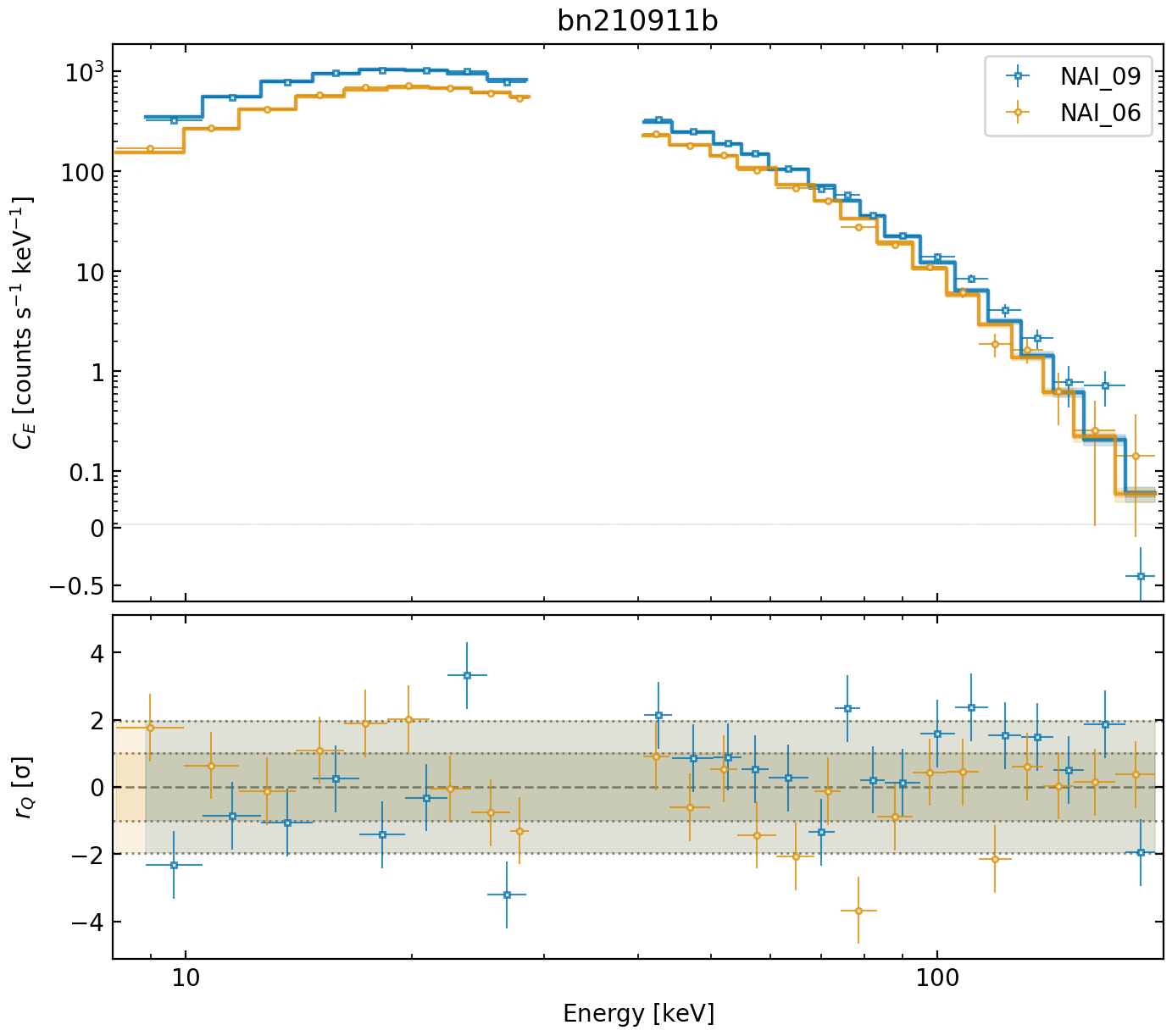}
\figsetgrpend

\figsetgrpstart
\figsetgrpnum{7}
\figsetgrptitle{bn210911c}
\figsetplot{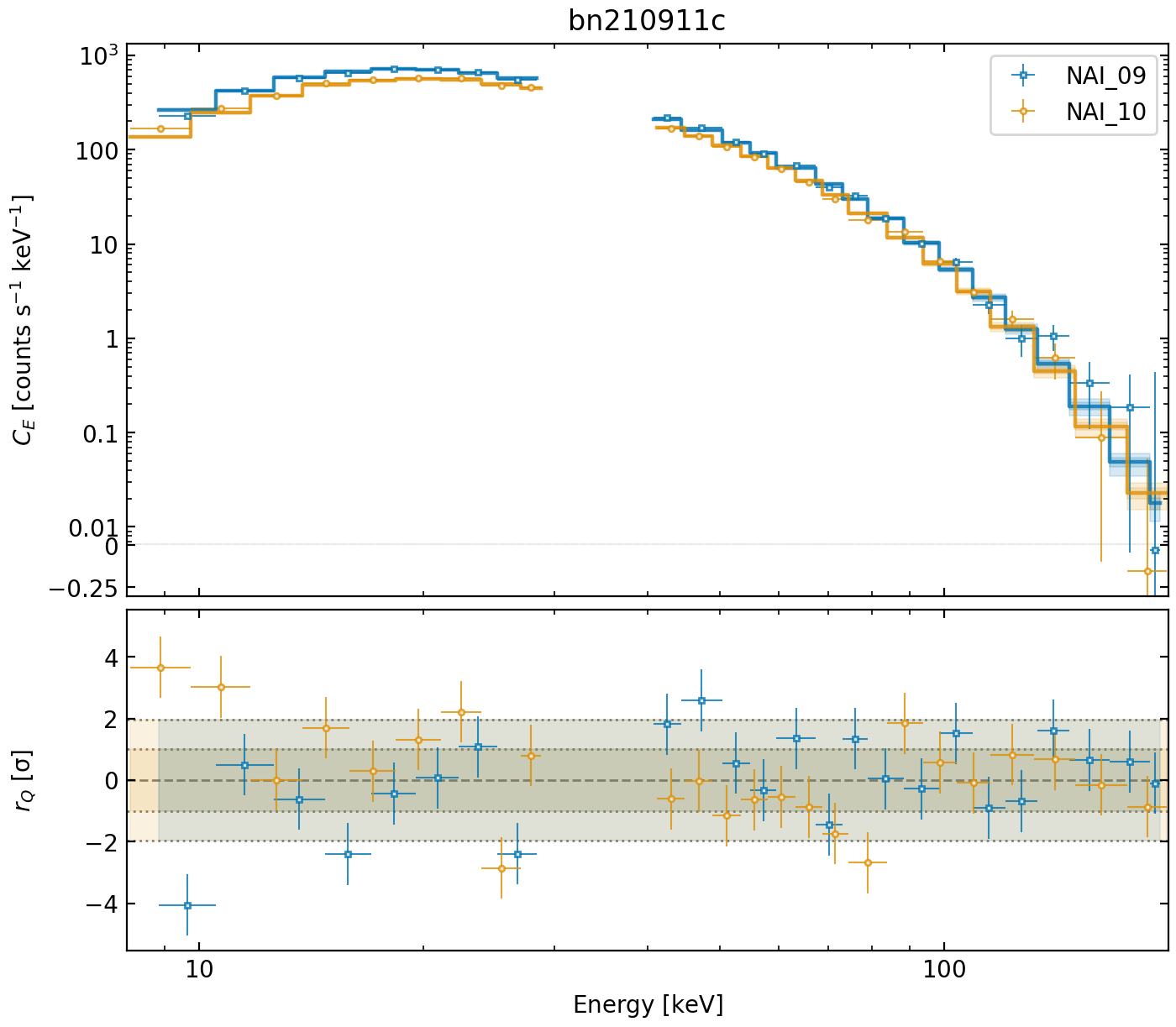}
\figsetgrpend

\figsetgrpstart
\figsetgrpnum{8}
\figsetgrptitle{bn211224}
\figsetplot{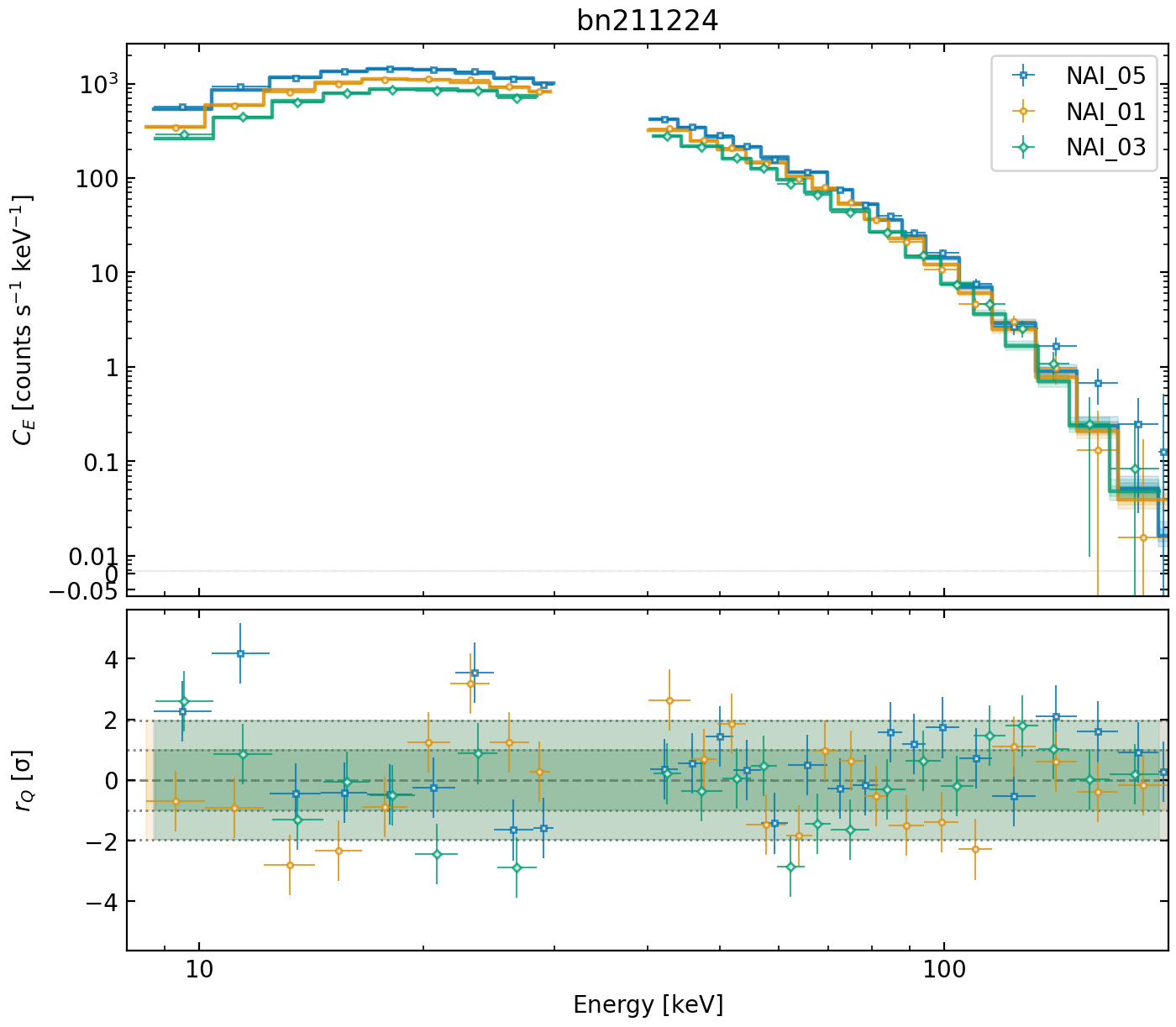}
\figsetgrpend

\figsetgrpstart
\figsetgrpnum{9}
\figsetgrptitle{bn220112}
\figsetplot{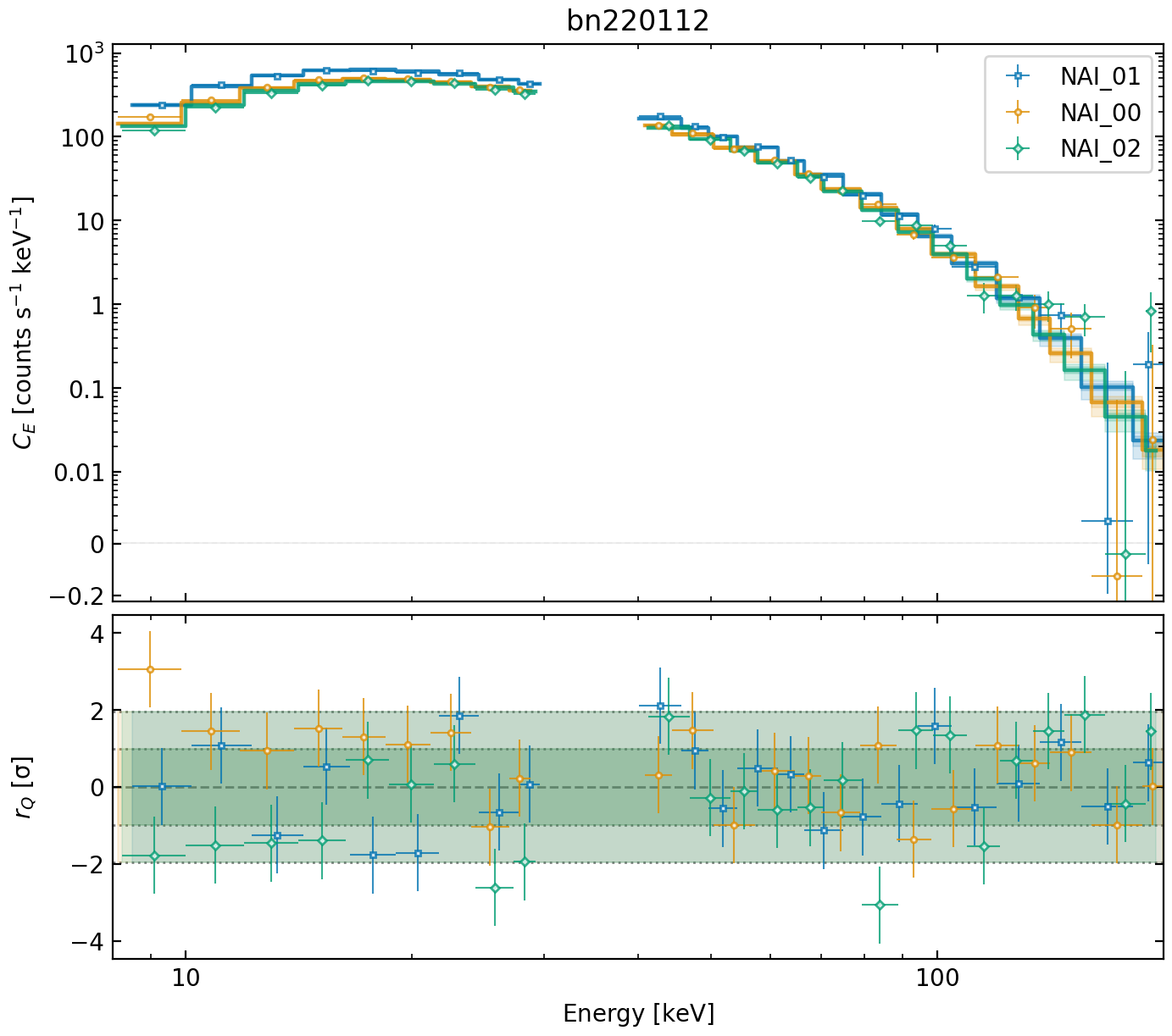}
\figsetgrpend

\figsetgrpstart
\figsetgrpnum{10}
\figsetgrptitle{bn210130}
\figsetplot{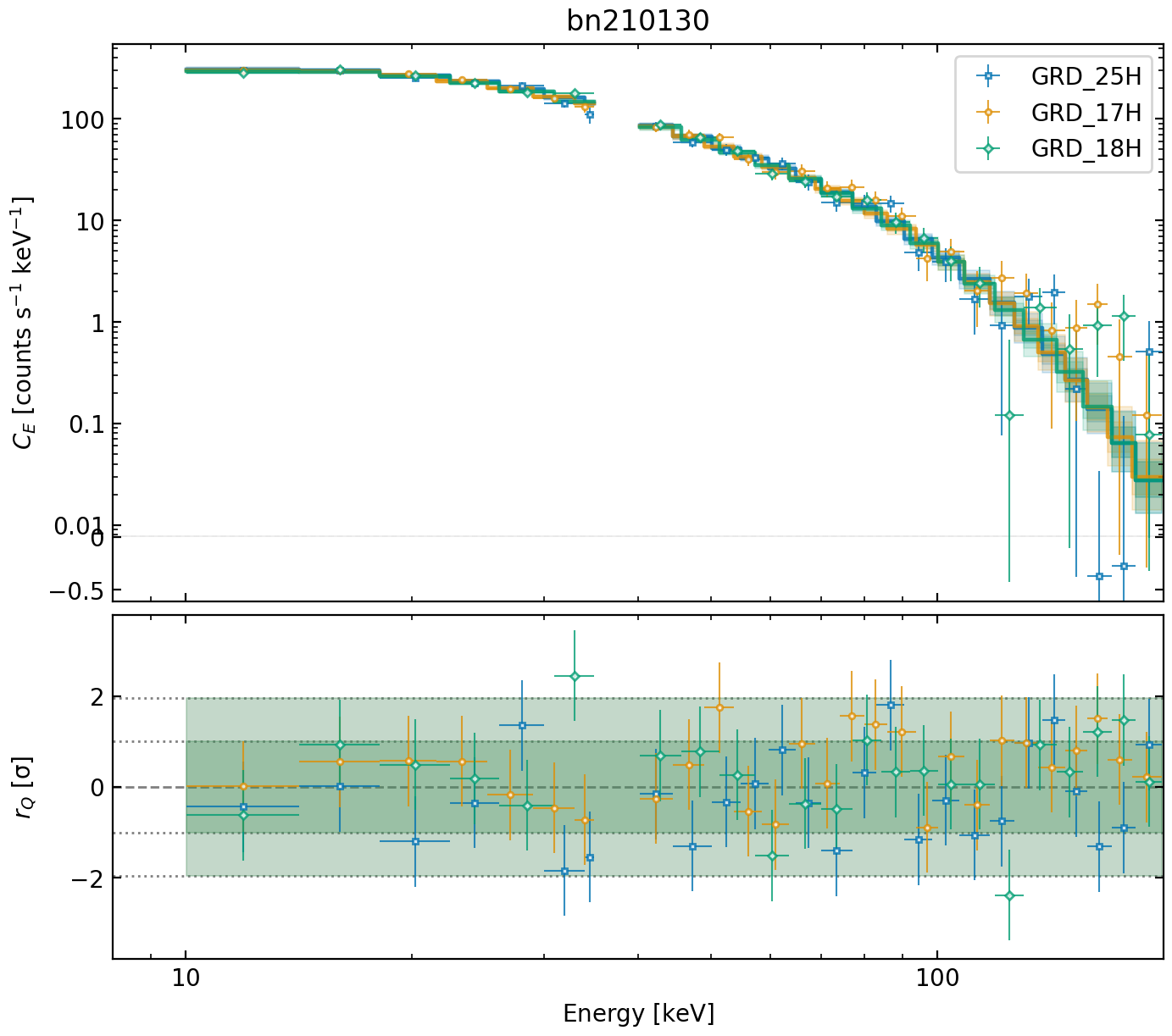}
\figsetgrpend

\figsetgrpstart
\figsetgrpnum{11}
\figsetgrptitle{bn210216}
\figsetplot{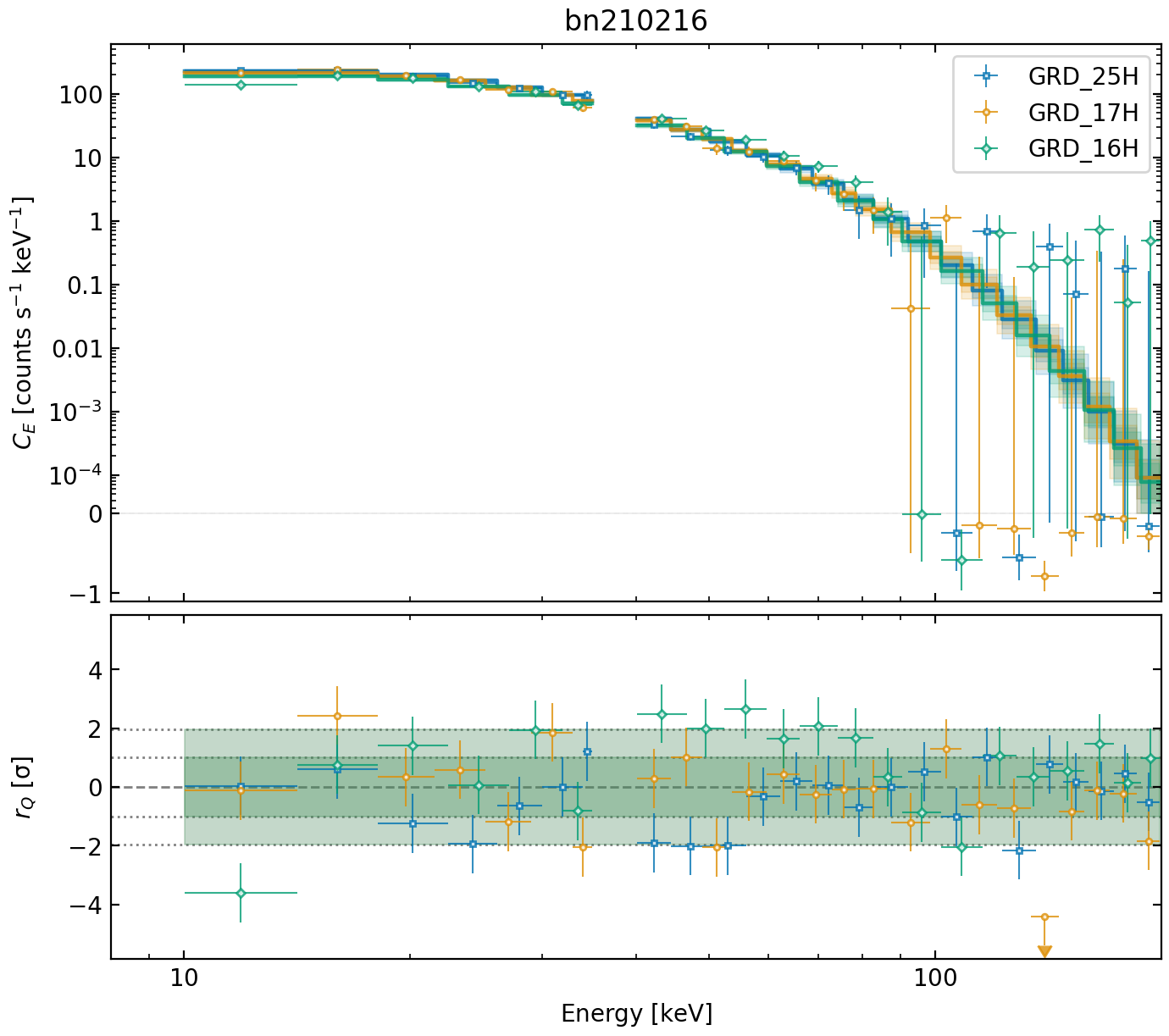}
\figsetgrpend

\figsetend

\end{CJK*}
\end{document}